\documentclass[acmsmall]{acmart}
\AtBeginDocument{%
  \providecommand\BibTeX{{%
    \normalfont B\kern-0.5em{\scshape i\kern-0.25em b}\kern-0.8em\TeX}}}

\setcopyright{acmlicensed}
\acmJournal{PACMHCI}
\acmYear{2024} \acmVolume{8} \acmNumber{CSCW2} 
\acmArticle{417} \acmMonth{11}\acmDOI{10.1145/3686956}

\usepackage{enumitem}
\usepackage{array}
\usepackage[utf8]{inputenc}
\usepackage{fontenc}
\usepackage{multirow}
\usepackage{booktabs}
\usepackage{tabularx}

\begin{document}

\title{Transitioning Together: Collaborative Work in Adolescent Chronic Illness Management}

\author{Rachael Zehrung}
\email{rzehrung@uci.edu}
\orcid{0000-0003-1617-9079}
\affiliation{%
  \institution{University of California, Irvine}
  \city{Irvine}
  \state{California}
  \country{USA}
}

\author{Madhu Reddy}
\email{mcreddy@uci.edu}
\orcid{0000-0002-6330-3637}
\affiliation{%
  \institution{University of California, Irvine}
  \city{Irvine}
  \state{California}
  \country{USA}
}

\author{Yunan Chen}
\email{yunanc@ics.uci.edu}
\orcid{0000-0003-4056-3820}
\affiliation{%
  \institution{University of California, Irvine}
  \city{Irvine}
  \state{California}
  \country{USA}
}

\renewcommand{\shortauthors}{Rachael Zehrung, Madhu Reddy, and Yunan Chen}
\renewcommand{\shorttitle}{Collaborative Work in Adolescent Chronic Illness Management}

\begin{abstract}
Adolescents with chronic illnesses need to learn self-management skills in preparation for the transition from pediatric to adult healthcare, which is associated with negative health outcomes for youth. However, few studies have explored how adolescents in a pre-transition stage practice self-management and collaborative management with their parents. Through interviews with 15 adolescents (aged 15-17), we found that adolescents managed mundane self-care tasks and experimented with lifestyle changes to be more independent, which sometimes conflicted with their parents’ efforts to ensure their safety. Adolescents and their parents also performed shared activities that provided adolescents with the opportunity to learn and practice self-management skills. Based on our findings, we discuss considerations for technology design to facilitate transition and promote parent-adolescent collaboration in light of these tensions.
\end{abstract}




\received{July 2023}
\received[revised]{January 2024}
\received[accepted]{March 2024}

\maketitle

\section{Introduction}
Chronic diseases are broadly defined as health conditions that interfere with daily life or require ongoing medical attention for a year or longer~\cite{cdc_about_2022}. Since chronic illnesses cannot be cured, chronic care management aims to enhance patients' quality of life and promote self-management to mitigate the negative impacts of illness. Managing chronic illness can be challenging due to the complexities of coordinating care~\cite{abowd_2006_challenges}, maintaining life-long treatment ~\cite{brown_medication_2011,chauke_factors_2022}, and coping with the effects of illness on daily life~\cite{turner_emotional_2000}. Personal informatics tools, which facilitate the collection and reflection of personal data~\cite{li_stage-based_2010}, have been shown to support chronic illness management by enabling patients to better understand and modify their health behaviors~\cite{chopra_living_2021, mamykina_investigating_2006}, as well as communicate more effectively with providers~\cite{chung_boundary_2016, pichon_divided_2021}. 

Chronic illness management is particularly challenging in childhood, as children often struggle to understand their conditions~\cite{beacham_children_2015}, adhere to treatment plans~\cite{dean_systematic_2010}, and cope with the emotional and psychological impact of chronic illness~\cite{barlow_psychosocial_2006}. Moreover, managing chronic illness in childhood requires extensive coordination and collaboration among children, their caregivers, and healthcare providers, which can strain family relationships~\cite{amir_care_2015, nikkhah_family_2022, seo_challenges_2021}. As chronic illnesses become more prevalent among youth~\cite{cdc_managing_2022}, it is increasingly important to understand how to support children's current health management practices and prepare them for future self-management. Prior work has examined how to facilitate self-management for children with chronic illnesses~\cite{shin_towards_2019}, as well as how parents and children collaborate to manage children’s health~\cite{cha_transitioning_2022, pina_dreamcatcher_2020, shin_more_2022}. However, much of this work takes a ``one size fits all'' approach (e.g., studying children aged 6 to 18)~\cite{su2024data}, while comparatively less work examines middle adolescence, defined as 15 to 17 years old~\cite{aap2017bright}.

Self-management skills are critical for adolescents with chronic illnesses as they prepare for their adulthood. Adolescents occupy a unique developmental stage in which they have increased capacity for self-management but still rely on caregivers to share the responsibilities of illness management. At the age of 18, adolescents must navigate the transition from pediatric to adult healthcare~\cite{white_supporting_2018}, a process often associated with reduced medication and treatment adherence (e.g.,~\cite{garvey_health_2012}), increased acute care usage (e.g.,~\cite{brousseau_acute_2010}), and poor health outcomes. Despite these risks, the majority of youth with special health care needs do not receive guidance from their pediatric care providers or parents on planning for this transition ~\cite{white_supporting_2018}.  To facilitate the transition of care, the medical literature emphasizes the importance of adolescent self-management~\cite{lindsay_barriers_2011}, as well as the need for parents to support adolescents in building the knowledge and skills needed for illness management~\cite{gray_barriers_2018, white_supporting_2018}. Although researchers in Human-Computer Interaction (HCI) and Computer Supported Cooperative Work (CSCW) have explored ways to support the self-management practices of adolescents~\cite{hong_using_2020, hong_care_2016, kim_toward_2019}, little work specifically examines their practices and perspectives regarding healthcare transition. 

To better understand adolescents' experiences, challenges, and perspectives related to the transition process, we conducted a qualitative interview study with 15 adolescents (aged 15-17) with a variety of common chronic illnesses. At this stage of life, adolescents exhibit greater independence from their parents and think more about future plans, making it an ideal time to study their health management practices before the transition to adult care. In this pre-transition phase, we found that participants and their parents performed separate yet complementary health management practices, which sometimes created tension between self-experimentation and parental oversight, as well as between privacy and safety. While participants managed mundane~\cite{nunes_understanding_2018} health tasks and sought knowledge of their conditions, their parents managed medical aspects of care and prioritized ensuring their safety. As part of shared illness management, participants engaged in joint activities such as shared note-taking and strategized about what health-related information to share with their parents. 

\vspace{5pt}\noindent{}The key contributions of this work include:
\begin{itemize}[leftmargin=*, itemsep=4pt, topsep=0pt]
    \item An empirical understanding of the needs and experiences of adolescents with chronic illnesses in a pre-transition stage. Adolescents felt motivated and prepared to manage their conditions independently through routine self-care but relied on their parents to handle the medical aspects of care and ensure their safety.
    \item A discussion of the tensions inherent in the transition process and their implications for collaborative illness management. Adolescents desired to self-experiment with their health but were concerned about their parents' oversight, leading to reduced sharing of health-related information. Similarly, adolescents sought control over their personal health information whereas their parents wanted to share information with peers and teachers in case of emergency. 
    \item Design recommendations for technologies that facilitate adolescents' participation in their healthcare and foster parent-adolescent collaboration, taking into account adolescents' values. 
\end{itemize}

\section{Related Work}
Our study is motivated by the lack of research exploring adolescents' health management practices and perspectives in the critical period before their transition from pediatric to adult care, as well as the potential for technology to support adolescents in becoming more active stakeholders in their health care. We begin by discussing chronic condition self-management and collaborative management in families, followed by a review of prior work on chronic illness in childhood and adolescence.
\subsection{Chronic Condition Self-Management}
Chronic illness self-management describes the day-to-day tasks that individuals undertake to manage their symptoms and reduce the impact of disease on everyday life~\cite{barlow_self-management_2002}. Designing technologies to support the self-management of chronic conditions has long been a focus of personal informatics~\cite{nunes_understanding_2018,schroeder_examining_2020}, and prior research has examined the management of various conditions such as diabetes~\cite{desai_personal_2019,mamykina_personal_2017,mamykina_investigating_2006}, asthma~\cite{davis_kiss_2018,lee_asthmon_2010,tinschert_potential_2017}, and migraine~\cite{park_individual_2015,schroeder_examining_2018}. A common approach to managing chronic illnesses is through self-tracking, which Li et al.~\cite{li_stage-based_2010} defined in their stage-based model as preparation, collection, integration, reflection, and action. Self-tracking technologies for chronic illness support the collection and review of patient-generated health data, typically with the goal of providing patients with actionable, data-driven insights to manage their symptoms and improve their well-being. For example, Geurts et al. proposed WalkWithMe~\cite{geurts_walkwithme_2019}, a mobile application that supports people with Multiple Sclerosis in setting goals and tracking their walking activity. Likewise, Karkar et al. introduced TummyTrials~\cite{karkar_tummytrials_2017}, a tool to support individuals with irritable bowel syndrome in identifying their personal food triggers through self-experimentation. While these studies demonstrate the potential for self-tracking technologies to increase patients' self-awareness and promote health behavior change, self-tracking can also contribute to overall treatment burden~\cite{ancker_you_2015,whitman_bodily_2021} and elicit negative emotional reactions~\cite{kabir_meta-synthesis_2023}. Self-tracking can be particularly frustrating for individuals with enigmatic chronic conditions such as Polycystic Ovary Syndrome and endometriosis due to heterogeneous symptoms and unexplained differences in response to treatment~\cite{chopra_living_2021,mckillop_designing_2018}.
Recognizing that self-tracking is a dynamic process, researchers have examined how to support the social and emotional needs of individuals living with chronic illness beyond medical treatment. For example, Ayobi et al.~\cite{ayobi_flexible_2018} highlighted the need for self-tracking technologies to support mental well-being and foster mindful experiences, rather than focusing solely on measurable disease indicators. In subsequent work, Ayobi et al. introduced Trackly~\cite{ayobi_trackly_2020}, a tool designed to facilitate meaningful tracking for multiple sclerosis, with a focus on supporting individuals' sense of agency. Nunes et al.~\cite{nunes_understanding_2018} drew a distinction between medicalized and mundane self-care technologies, arguing that the former are designed to extend medical treatment to the home rather than support the complexities of everyday life with a chronic illness. In examining the everyday activities performed by individuals with Parkinson's, they found that self-care was achieved in mundane (i.e., routine or practical) ways that were shaped by individuals' competing concerns. We adapt Nunes et al.'s conceptualization of mundane self-care to describe adolescents' management practices, which were similarly focused on routine activities that prioritized daily life and well-being over disease monitoring. Our findings add to an understanding of the factors that shape mundane self-care by highlighting how adolescents' self-care practices were affected by collaboration and their relationships with others. 

\subsection{Collaborative Illness Management in Families}
Given that adolescents in the pre-transition phase (i.e., under 18) typically live with their families and rely on their parents to share responsibility for illness management, it is essential to consider the role of family members in adolescent health management. While much of the personal health informatics literature focuses on individual tracking needs, the field has begun to explore the benefits of collaborative tracking within clinical care teams and families~\cite{epstein_mapping_2020}. Collaborative health management can help family members understand how their individual behaviors and health are interconnected~\cite{pina_personal_2017}, as well as facilitate deeper reflection on individually collected health data~\cite{grimes_toward_2009}. Additionally, establishing family goals can promote health discussions and supportive behavior within families~\cite{colineau_motivating_2011}. 
Prior work on collaborative illness management in families tends to focus on the experiences of caregivers~\cite{bhat_we_2023,jacobs_i_2019,nikkhah_family_2022} and young children~\cite{jo_mamas_2020,shin_more_2022}. However, as children reach adolescence, their needs and preferences shift. For example, Toscos et al.~\cite{toscos_best_2012} found that health monitoring technologies could exacerbate tensions between adolescents with Type 1 Diabetes and their parents. While adolescents desired independence, their parents experienced anxiety and a lack of trust in adolescents' self-management abilities. Though younger children may benefit from close health monitoring, collaborative management with adolescents requires a different approach due to their growing desire and capacity for self-management. Recognizing this need, Hong et al. investigated ways to engage adolescents in their care~\cite{hong_adolescent_2017} and support gradually evolving partnerships with caregivers~\cite{hong_care_2016}. In another study, Hong et al.~\cite{hong_using_2020} demonstrated how technology can support shared illness management with adolescents and their parents by scaffolding collaborative reconstructions of illness experiences. Motivated by these studies on the benefits of collaborative illness management and the unique needs of adolescents, our study explores adolescents' relationships with their parents and how shared practices can support adolescent health. In contrast to other work, our study examines adolescents' attitudes towards transition, how they learn and practice self-management to prepare for transition, and which collaborative  practices help or hinder their transition. 

\subsection{Chronic Illness in Childhood and Adolescence}
Compared to their healthy peers, youth with chronic illnesses report lower levels of emotional well-being and greater difficulty in establishing independence from their parents~\cite{yeo_chronic_2005}. Recognizing the importance of fostering children's autonomy~\cite{kawas_another_2020}, researchers have proposed systems to help children better understand and engage in health management (e.g.,~\cite{ankrah_me_2022,lee_asthmon_2010,shin_towards_2019}). In their study on how parents and children collaborate to manage Type 1 Diabetes, Cha et al.~\cite{cha_transitioning_2022} found that children's knowledge of illness management and motivation for self-care were critical in their transition towards independence. While these studies underscore the need to support youth self-management and engagement, they primarily focus on children under the age of 15, whose needs and preferences differ from those of adolescents. 
During adolescence, parent-child relationships evolve, becoming more interdependent and reciprocal as adolescents strive for autonomy and reduced parental control~\cite{branje_development_2018}, which can lead to conflict~\cite{babler_moving_2015}. This developmental period is critical because it precedes the transition from pediatric to adult care, which is associated with a drop in treatment adherence and negative health outcomes for adolescents~\cite{gray_barriers_2018,lindsay_barriers_2011}. A successful transition also requires interdisciplinary care coordination between adolescents, family members, and pediatric and adult health care providers~\cite{lemke_perceptions_2018}. While prior work has focused on parents' views on transitional healthcare for their children with chronic illnesses~\cite{heath_parenting_2017,lebrun-harris_transition_2018}, fewer studies have examined adolescents' experiences and views regarding their own transition~\cite{lerch_adolescents_2019}. Heath et al.~\cite{heath_parenting_2017} found that parents tried to prepare adolescents for healthcare transition by monitoring condition management, prompting treatment adherence, and offering technical assistance. Although parents might view these strategies as effective, it is important to understand adolescents' perspectives on which activities they find helpful to their transition. Further, while existing research highlights the importance of shared responsibility between adolescents and parents, the day-to-day activities that shape parent-adolescent collaboration have not been well studied. Our work sheds light on adolescents' everyday experiences, challenges, and health management practices, as well as their attitudes towards the transition to adult care.

\section{Methods}
To investigate the self-management and collaborative management practices of adolescents with chronic illnesses, we conducted semi-structured interviews with adolescents diagnosed with a variety of chronic conditions. The interviews provided rich data around their lived experiences, needs, and challenges. This study was approved by the university's Institutional Review Board. 

\subsection{Participants}
We conducted 15 semi-structured interviews with adolescents (aged 15-17), all of whom had been diagnosed with a chronic condition (see Table~\ref{tab:my-table} for participants' characteristics). Unlike prior work, which often focuses on populations with specific illnesses (e.g., diabetes and asthma), we sought to uncover the common experiences and challenges shared by adolescents with different chronic illnesses as they navigated this transitional period, with a focus on their interactions with parents, doctors, and peers in health-related contexts. Participants were recruited through word-of-mouth, snowball sampling~\cite{parker_2019_snowball}, and online distribution of the study flyer (e.g., on Reddit and Twitter). Recruitment occurred from July 2022 to April 2023. Eligibility criteria included a diagnosis of a chronic condition and being between the ages of 13 and 17. 
Interested individuals were invited to complete an online screening questionnaire that asked if they had been diagnosed with a chronic condition (yes/ no), their diagnosis, and their email address, as well as a parent's email if the respondent was under the age of 18. We followed up with eligible respondents via email and invited them to participate in the study following completion of the assent form. We also reached out to parents to seek their permission for their children's participation and asked them to complete the consent form. On the assent form and before the interview, participants were asked if they wanted to allow both audio- and video-recording or only audio-recording. 


\begin{table}[]
\resizebox{\textwidth}{!}{%
\begin{tabular}{@{}lcccc@{}}
\toprule
\textbf{Participant} & \textbf{Age} & \textbf{Diagnosis} & \textbf{Gender} & \textbf{Family Structure} \\ \midrule
Mark & 17 & Asthma & Male & Two-parent, siblings \\
Elijah & 17 & Complex Post Traumatic Stress Disorder & Male & Single-parent (father) \\
Maya & 15 & Type 1 Diabetes & Female & Single-parent (mother), sibling \\
Naomi & 16 & Asthma & Female & Single-parent (mother), sibling \\
Hana & 17 & Polycystic Ovary Syndrome & Female & Two-parent, sibling \\
David & 17 & Type 1 Diabetes & Male & Separated, sibling \\
Jon & 17 & Asthma & Male & Two-parent, siblings \\
Ben & 16 & Type 2 Diabetes & Male & Two-parent \\
Liam & 17 & Type 1 Diabetes & Male & Single-parent (father) \\
Alisha & 17 & Chronic Kidney Disease & Female & Two-parent, siblings \\
Aaron & 17 & Asthma & Male & Single-parent (father), siblings \\
Nia & 15 & Asthma & Female & Single-parent (mother), siblings \\
Sara & 15 & Asthma & Female & Two-parent, siblings \\
Emma & 16 & Asthma & Female & Two-parent, siblings \\
Simone & 17 & Chronic Fatigue Syndrome & Female & Two-parent \\ \bottomrule
\end{tabular}%
}
\caption{An overview of participants’ characteristics. All participants have been referred to with a pseudonym. Seven participants lived in two-parent households; seven participants lived in single-parent households; and one participant split time between households. The most common diagnoses were asthma and diabetes.}
\label{tab:my-table}
\end{table}

\subsection{Data Collection}
Interviews were conducted online and lasted up to an hour using a semi-structured approach with follow-up questions when necessary. We started the interviews by asking participants about their health history, family dynamic, and the impact of their condition on their personal lives. Participants were then asked to describe how they manage their condition individually and collaboratively with their parent(s) and other family members. Lastly, participants were asked about their technology use in relation to health management (e.g., tracking tools and social media use). All interviews were conducted over Zoom, recorded, and transcribed using Zoom's built-in automated transcription service followed by manual error correction. After transcription, the interview recordings were deleted and transcripts were de-identified. 

\subsection{Data Analysis}
We employed thematic analysis~\cite{braun_thematic_2012} to qualitatively analyze the interview data. Using open coding, the first author analyzed two transcripts and generated an initial set of codes. The initial codes were diverse and included topics such as adolescents' experiences with stigma and peer interactions. We observed particularly rich data around how adolescents and parents collaborate around adolescents' health. Consequently, we re-coded the data with a focus on health management practices. After several rounds of discussion, we performed axial coding~\cite{williams_2019_art}, generating new codes when necessary. Following an iterative approach, we organized the codes according to overarching themes and related interview quotes to one or more themes. For example, ``shared note-taking with parents'' and ``going with parents to the doctor's office'' were organized under the theme, ``parental support for medical management tasks.'' The final codebook consisted of 8 themes and 25 codes. 

\subsection{Ethical Considerations}
Through our interviews, we aimed to understand adolescents' perspectives and experiences around health management in both personal and family contexts. Given that adolescents are a vulnerable population, we were mindful in our approach to sensitive questions. Participants were informed in the assent form and again before the interview about their right to refuse to answer any question or to stop the interview at any time. We reassured participants that they would not face any consequences for skipping questions or ending the interview. All participants completed the interviews, with several reflecting that the conversation was beneficial (e.g., ``\textit{thank you for the interview. It just helped me like relieve some of the thoughts I have}'' -- Maya). Additionally, we sought parents' consent for their children's participation and encouraged them to discuss the study by requiring both the parent and child to sign the consent form. In the paper, we also omit certain details (e.g., names of support groups) to protect adolescents' safe spaces. 
\section{Findings}
In this transitional stage, we found that adolescents and parents performed distinct yet complementary aspects of health management. Adolescents handled mundane day-to-day health management tasks and attempted to manage their conditions through lifestyle changes, while they perceived their parents as primarily focused on ensuring their safety and teaching them illness management skills. These differing roles sometimes led to tension, as adolescents wanted more autonomy to self-experiment and exercise control over their personal lives, whereas they believed their parents wanted to oversee their activities and stay informed about their health. We also highlight shared management practices typical of this developmental stage, where adolescents have a growing desire and capacity for self-management but still rely on their parents to help shoulder the responsibilities of illness management.

\subsection{Adolescents: Managing the Mundane}
Participants were motivated to learn about their conditions independently and assume more responsibility for mundane, everyday health management tasks such as taking medication. As 15 to 17 year olds, most participants were able to access information online, which they often used to inform their coping strategies and implement lifestyle changes. They also sought to build connections outside of their families by joining online support groups and learning from other individuals with the same condition. 

\subsubsection{Learning and Practicing Self-Care}
Most participants expressed a strong desire to manage routine health tasks independently and felt confident in their ability to do so. They first looked to their parents to learn about health management and then supplemented this knowledge through online web searches and social media. Ben, for example, learned how to check his glucose levels by observing his parents: \textit{``[My parents] always did it for me since the age of 12},'' but \textit{``I started checking it by myself in the past year because I told them I could check it myself because I always see how they do it.''} Some participants who were diagnosed at a younger age felt more confident about their health management skills after observing their parents' actions since diagnosis. In this case, Ben asserted his ability to independently check his glucose levels four years after being diagnosed with diabetes, demonstrating a clear shift in responsibility for illness management. 

However, not all participants felt supported by their parents in their efforts to become more independent. Maya expressed, \textit{``I really want to be independent... so I really want to be able to learn about blood sugars.''} Although Maya \textit{``tried to ask about it},'' her mother told her that measuring her sugar is \textit{``complicated''} and that she will \textit{``just learn in the future.''} For a few participants, their parents' unwillingness to teach them specific skills led to a loss of motivation and knowledge deficit. As David explained, \textit{``it’s kind of medical stuff. I don’t really understand it. I'm seeing T1, T2, sugar levels and other stuff.''}  Though he sometimes went online to \textit{``search for other stuff to do to lessen sugar},'' he admitted that \textit{``95\%, ok, let me say 90\% [of information] is from my dad.''} Adolescents such as David did not understand the medical complexities of their conditions and instead viewed health management as following sets of instructions from their parents and doctors, which hindered their movement towards independence.  

More commonly, when adolescents encountered these barriers to medical understanding or resistance from their parents, they explored other means to take more control of their health. The majority of participants actively sought information related to their condition through social media and web searches to learn how to manage their health through lifestyle changes. Simone shared, \textit{``if you're trying to like, really understand your condition... [YouTube has] so many videos about [Chronic Fatigue Syndrome] you watch and suggestions about what types of activities you can do, or like yoga, meditations.''} Similarly, Hana turned to \textit{``dietician TikTok''} to learn healthy recipes that met her food restrictions, because \textit{``that’s my biggest problem, it's food.''} While participants were most comfortable with consuming content from social media, some engaged with creators and other viewers to exchange management tips based on real-life experiences. For example, Naomi turned to TikTok to \textit{``watch DIY videos on things to do to remedy some situations''} and connect with people who are \textit{``passionate about what's going on in other people's lives, and they always want to help out if you have a situation or problem. They try to make suggestions, recommend new stuff, and I really like that.''} While participants appreciated social media for suggestions and sometimes personal connections, they also used simple web searches to find answers to specific questions or issues. Jon shared, \textit{``I just Google everything. If I hear something I don't understand... while they're talking about it in the hospital, I can just Google it and make sure I understand.''} Simone used a chatbot for a similar purpose: \textit{``if I want to get information, I pretty much just use myAI... I can ask it like-- question then suggestion... it's like, more personal than Google.''}

\subsubsection{Health Data Tracking}
Other than learning online, participants learned more about their conditions by tracking or recording their health data, which allowed them to identify patterns in their symptoms and adjust their lifestyles. As one participant with asthma described, \textit{``over the years, I've been trying to study which days my symptoms might show up... I took a lot of notes down. I wrote dates; I wrote times; I wrote some medication''} (Mark). Like many participants, Mark recorded his data manually in physical diaries or notepads, as opposed to specific tracking applications. Part of mundane self-care is making lifestyle adaptations, which participants were able to do by tracking their health and developing a greater sense of awareness around their triggers and the impact of lifestyle factors. After tracking his symptoms, Mark set reminders \textit{``that could tell me what time I need my inhaler, what time I should drink water.''} Based on their observations, participants sought to control their environments and activities to avoid triggers and protect their well-being. For example, Elijah understood that certain social contexts (e.g., crowds) would worsen his CPTSD symptoms: \textit{``the most I'll do is just move away from that environment, like maybe instead of being at the game, I just head to class.''} Participants' home environments were relatively controlled, as their parents took care of household tasks such as cooking and cleaning. Outside of the home, however, participants had to learn how to cope with their environment. With their increased autonomy, participants spent more time outside of the home and without their parents, which resulted in increased decision-making opportunities related to health matters. 

Although many participants found value in tracking, some were hesitant to record their data due to perceived tension between self-experimentation and parents' safety concerns. Jon shared that he practiced running after school hours \textit{``to see how long it takes me to run before I start feeling that uneasiness in breathing.''} Though he had \textit{``seen a lot of improvement},'' he only kept \textit{``mental track''} of his results because of \textit{``the whole security thing''}: \textit{``I don't want anyone to find out... [my parents] would just tell me to stop instantaneously''} (Jon, asthma). Though self-experimentation is common in self-tracking, participants felt that they could not be open about some of their health behaviors and experimentation because of their parents' safety concerns. Parental support (or lack thereof) could be a deciding factor in whether adolescents' engage in self-tracking and record-keeping. As Jon explained, \textit{``if my parents were to support me and give me the go ahead, I would definitely love to keep records.''}

\subsubsection{Managing Relationships through Data Sharing}
Beyond the tension over safety concerns, many participants did not necessarily want their parents to have access to all their data because they associated ownership of their data with independence. For example, Aaron tracked a lot of data but did not feel comfortable sharing everything with his parents: \textit{``I share almost everything. But with that, I still kind of hold back a little bit. Maybe like 80\% is what I share... I'm trying to be an open book, but right now I'm an adult.''} Aaron previously shared everything with his parents but felt less inclined to share data once he perceived that he was fully an adult, reflecting a broader trend among participants whose perspectives on appropriate boundaries with their parents shifted as they gained independence. In deciding whether to share, participants also weighed the perceived health benefit with the impact that sharing might have on their relationship with their parents. 
Some participants shared everything health-related with their parents because they believed it would improve their well-being. Maya shared health updates every evening with her mother because she believed that withholding information could \textit{``harm [her] in the near future},'' and \textit{``sharing everything about [her] symptoms''} would allow her to \textit{``really get better and be healthy.''} As an example, Maya pointed out that sharing information about a new or worsened symptom with her mother could lead them to visit the doctor's office to modify her treatment plan. Participants saw value in sharing information with their parents as a way to monitor their symptoms and better manage their conditions, and like Maya, many of them shared everything health-related because they believed it was in their best interests. These health updates typically took place during routine interactions  such as \textit{``after homework''} (Nia), \textit{``over dinner... during storytime''} (Jon), and \textit{``during breakfast''} (Simone), providing participants with consistent opportunities to share their concerns, reflect on their management, and plan any necessary changes to their lifestyle or treatment.

In contrast to providing daily health updates and sharing everything, a few participants approached information-sharing on a need-to-know basis due to the perceived impact of sharing on their relationships with their parents. Hana, for example, did not share health-related information with her parents \textit{``unless something’s really important.''} In this case, the participant avoided telling her parents about her health because \textit{``my mom gets very upset about things, like she stresses a lot},'' which she wanted to avoid. Adolescents like Hana were aware of how sharing health-related information might impact their parents, and some withheld information to lessen their parents' hardships. As Naomi shared, \textit{``most times I just don't want to disturb her with my problems, because I feel she’s going through a lot being a single mom. So most times, I try to... improve myself, so like self-care and all.''} Similarly, Sara did not talk to her parents about her health often because \textit{``I really don't want to burden my mom''} and her dad was stressed with managing \textit{``everything at home.''} These participants were attuned to their parents' struggles and did not want to cause more worry by bringing up their health, unless an emergency or something out of the ordinary happened. 

\subsubsection{Finding Community}
While participants benefited from online support groups that had members of all ages who were diagnosed with the same condition, they found particular value in connecting with others around the same age: \textit{``I feel supported in this talking with people of my age mostly, because we are experiencing so many things which are similar... they are going through the same thing, and they are also at school''} (Nia). Peer groups become increasingly important during adolescence, while familial influence declines, which can be challenging for adolescents with chronic illnesses as they struggle to find a sense of normalcy and belonging with their healthy peers. In our study, participants actively sought online support groups to find this sense of belonging and learn from other individuals' lived experiences. For example, Maya joined several Facebook groups for individuals with diabetes to see \textit{``suggestions on how you can control the situation and the meals you should do},'' especially because she did not know anyone else in her personal life with diabetes. Similarly, Aaron joined a group for individuals with asthma, which he described as \textit{``a community where we come up with ways on how we can better our lives''} that allowed him to learn from \textit{``the challenges that the rest are facing out there, and how they are coping with them.''} By connecting with other people with the same diagnosis, participants were able to find a sense of community, establish a sense of normalcy, and learn how to better manage their conditions. While most participants found these groups on their own, a few were recommended by their doctors to try online support groups. Emma was added to a group message for youth with asthma by her doctor and shared that through the group, \textit{``I've learned a lot about asthma and how to manage it... They're so loving, caring. And they really check up on someone, like if you get an attack, and then you tell them, they will tell you it's going to be over; it's just a season.''}

\subsection{Parental Support: Ensuring Safety}
Although participants felt confident about managing their daily routines, they were not necessarily prepared to handle other aspects of health management. They relied upon parents to supervise self-care tasks and mediate social interactions both within and outside of the family. While adolescents' primary responsibilities focused on their day-to-day routines, their parents' role was mainly to ensure their safety. However, parents' efforts to ensure safety sometimes caused tension, as adolescents felt their personal boundaries were not respected. Importantly, these findings are based on adolescents' perceptions of their parents' behaviors and actions, as we did not interview parents themselves.  

\subsubsection{Ensuring Safety at Home}
As our study participants described, their parents ensured safety at home by confirming or supervising the completion of medical tasks (e.g., taking medication) and splitting responsibilities among family members, including siblings. One participant shared, \textit{``most of the time [my father] would ask me, have you taken your meds, make sure to take your meds. He would always say that even if I have already done so. He would just make sure''} (Elijah, CPTSD). In this case, Elijah was primarily responsible for taking his medication everyday, whereas his father's role was to remind him and double-check his adherence. Similarly, Ben checked his blood glucose levels by himself, but \textit{``sometimes my dad confirms it.''} These intermittent safety checks allowed participants to practice self-management and enabled parents to ensure their safety and long-term treatment adherence. After a diabetic emergency, Maya's mother started to directly supervise her medication regimen: \textit{``She loves to sit there and make sure that I'm taking the drugs... maybe she thinks that I'm not old enough to handle taking my medication.''} In this instance, the participant's mother reacted to her health emergency by becoming more restrictive and less trusting, as she worried that further deviations from treatment would result in negative health outcomes. Likewise, Hana's parents attempted to supervise or confirm her medication adherence, leading to tension: \textit{``[my mom] is just like, I don't think you've been doing all these things. Like, but I have, like I don't know how to convince you that I have.''} Participants often felt that they were capable of managing their health, but that their adherence was not recognized with a commensurate increase in trust or responsibility due to their parents' safety concerns.

Parents also split responsibilities and coordinated their schedules to ensure care coverage (e.g., \textit{``[taking] shifts''} -- Alisha) and develop contingency plans. In homes with siblings, parents often equipped both older and younger siblings with the knowledge and skills needed to assist with illness management and emergencies, particularly for when parents were not present. As one participant shared, \textit{``Mom made sure that everybody in the house has a spare [inhaler], and they are spread in separate locations... [my sisters] are always very vigilant, just like my parents''} (Jon). This participant's parents both worked long hours outside of the home and were concerned about the potential for asthma attacks while they were away. They involved the entire family in contingency planning by training his sisters to monitor him and distributing inhalers throughout the house for ease of access in case of emergencies. Similarly, Nia's younger sister was taught how to handle emergencies: \textit{``my mom told her about it, and also told her what she can do in case anything happens.''} For single-parent families like Nia's household, siblings were instrumental in providing an extra layer of safety planning for adolescents. 

\subsubsection{Involving Peers in Illness Management}
As our participants described, their parents often wanted to involve their peers as a way to check-in on their health outside of the home and ensure their safety in case of a medical emergency. To that end, parents encouraged participants to share their health-related information with peers, despite the potential discomfort or stigmatization associated with disclosure. For example, Naomi's mother encouraged her to share her asthma with her friends as a health precaution: \textit{``[My mom] actually knew that I was hiding stuff and I wasn't comfortable and ready to accept my condition. So she actually talked to me and advised me to go ahead, to tell my friends.''} Despite her reluctance to inform her friends, her mother \textit{``had to meet up with them, my close friends, and just tell them about it''} after the participant experienced another asthma attack. Here, Naomi's mother decided that it was more critical to ensure her safety than to follow her preference for non-disclosure. Following the attack, Naomi's mother equipped her friends with the skills and tools (i.e., inhalers) needed for medical interventions: \textit{``a few of them carry extra inhalers in their backpacks for me just in case I don't have mine.''} Peer disclosure and education mitigated the risk of peer pressure as well, because then participants' peers knew about the risks of their condition. For example, Maya experienced a diabetic emergency after her friends \textit{``really convinced [her] that eating it once in a while won't affect [her].''} While these risk-taking behaviors and peer influence are typical of adolescence, they can have more severe consequences for adolescents with chronic illnesses. As Naomi pointed out, her mother wanted her to share her condition \textit{``so [my friends] know how to treat me and... be careful around me.''} Similarly, another participant's father informed her best friend about her asthma and \textit{``actually taught him how to stay with me and how to handle everything''} (Sara). While the participant's father did not consult her about this disclosure, Sara appreciated his involvement because it \textit{``felt good seeing that he cared''} and resulted in her best friend visiting her more often to check up on her. As mediators, parents helped build participants' support networks, particularly when adolescents themselves felt uncomfortable or unable to reach out to other individuals for support. 

In addition to teaching participants' friends how to help in emergencies, some parents directly consulted with peers to receive health updates. For example, one participant shared that part of her mother's daily check-in routine is to contact her best friend: \textit{``She talks to her and asks her how I did at school or if there's any emergency that occurred''} (Maya, diabetes). As adolescents develop more autonomy and spend more time outside the home, parents have a shrinking amount of influence on their daily interactions. By calling her friend, the participant's mother was able to gain a fuller picture of what occurred during the school day while out of her purview. Parental disclosure to peers sometimes resulted in tension, as adolescents believed that their parents were overstepping their personal boundaries and not respecting their ability to self-manage. As Maya expressed, \textit{``I'm so, so uncomfortable with it because it's like she's exposing to the world that I have diabetes... she usually keeps on telling me that it's for my well-being, but I just don't feel that she should be calling to check on me each and every day... I know my condition, and I know that I need to keep myself well.''} Adolescents with chronic conditions lack complete control over when, how, and to whom they disclose their conditions. Some participants did not want to tell any of their peers about their illness, yet their ability to withhold personal information was compromised by their parents' decision to involve their peers in illness management. 

\subsubsection{Ensuring Safety at School}
 Participants relayed that their parents informed teachers of their conditions to request accommodations and monitor their health in case of emergency. For example, some participants' parents had to inform teachers to explain school absences and allow them to make-up the work. As Nia shared, \textit{``[my teachers] thought maybe it's an excuse because of not coming to school... so after that, my mom was requested to go to school so that she can explain to the teachers and bring some medical documents.''} Although Nia's mother extended the opportunity for her to disclose her condition to teachers, the participant's mother ultimately had to engage with teachers directly to provide evidence of the illness, which the participant could not do on her own. Even when parents want to encourage autonomy, adolescents often encounter barriers that they cannot overcome independently, necessitating parental support. That said, a strong parent-teacher relationship can benefit adolescents' well-being in a school context. Nia shared that after she experienced difficulties with her peers excluding her socially, her teachers \textit{``had a session''} in which they \textit{``explained why I'm asthmatic, what are my triggers, and what [my schoolmates] can help me do or take me to the nurse when I have those asthmatic attacks.''} This session was requested by the participant's mother \textit{``after I complained and she saw that I was distracted when I was at home.''} By meeting with her teachers beforehand, Nia's mother built a relationship with the participant’s teachers and was able to advocate for her needs at school, both medically and socially. 
 
According to participants, parents also disclosed adolescents' conditions to teachers to monitor their health and intervene in medical emergencies if necessary, which sometimes led to tension.  One participant described how his parents requested for his teacher \textit{``to pay close attention''} to him, which he found \textit{``really uncomfortable''} and \textit{``weird''}: \textit{``I didn't even know that [the teacher] was checking my vital signs... I would turn around, the whole class is just looking at me and everybody is wondering what's going on. I'm like, well, what could possibly be going through their heads?''} (Jon). Although Jon's parents did not directly disclose his condition to his peers, their disclosure to his teacher and the teacher's subsequent behavior exposed his condition to his classmates. When this routine check-up became too visible, some participants felt singled out from their peers and too closely monitored in an environment where they were accustomed to having more freedom from their parents. 

\subsection{Managing Together: Pathways Towards Independence}
The joint performance of health management activities allowed parents to teach or demonstrate important life skills, while providing opportunities for participants to learn or practice their management skills in a safe and controlled environment. In this transitional stage, parents gradually reduced their involvement in mundane, everyday care tasks and equipped participants with the skills needed to become more independent. 

\subsubsection{Communication and Reflection through Note-Taking}
Parents and participants communicated and learned together through shared note-taking around daily life.  Participants often found it challenging to articulate their thoughts and feelings, particularly around new symptoms and experiences, and writing notes enabled them to process their emotions, communicate more clearly with their parents, and reflect on their health. Participants and their parents engaged in shared note-taking in two ways. In one method, they both contributed and reflected on adolescents' health information; in the other, participants provided information but left the reflection to their parents. 

In many cases, parents and participants shared the responsibility for record-keeping evenly, and participants were heavily involved in interpreting and learning from their records. After her diagnosis, Alisha started a system with her parents to facilitate communication around her treatment: \textit{``There was this, like, cabinet in the house where you could drop sticky notes, and the sticky is not meant to be disposed of.''}  The cabinet was in the participant's room, giving her access to the notes at all times. She left notes for her parents \textit{``especially when I feel weird, or you know, I developed symptoms that I'm not so sure how to explain. I would always write it down to make them understand better.''} Alisha's parents would also leave notes to remind her of changes to her treatment plan (e.g., \textit{``this is a new medication I need to try''}) or to document her progress (e.g., \textit{``if there’s an improvement in the balance of my high blood pressure''}). The permanence of this installation enabled the participant and her parents to record symptoms over time. While Alisha and her parents contributed notes separately, it was more common for participants and parents to write notes together. As Maya stated, \textit{``we have a diary where we usually write the cases only when the conditions were to the extreme... we usually write like three to five things in a week.''} This participant and her mother would discuss her day after school, and they would add notes to a shared diary to keep track of her diabetes-related symptoms. This diary also served as a record for Maya's doctor, and she brought it with her to appointments, which was particularly useful when her mother could not go with her. Similarly, Emma would orally discuss her asthma attacks with her parents and answer questions (e.g., \textit{``what did you feel this time when the attack came''}), which they recorded for her (\textit{``in a notebook... they write the date the attack came and the symptoms''}). With this notebook, Emma said \textit{``we can compare from the previous attack''} at the end of every month, and that she has learned \textit{``how to feel the asthma attack coming... how to manage myself, not to use the inhaler.''} By asking guiding questions about each attack, Emma's parents encouraged her to think about her triggers and involved her in health management. By recording her symptoms, they provided her with a valuable record to facilitate further reflection about her asthma attacks and enabled her to learn how to predict and manage these attacks. 

In contrast, a few participants were responsible for providing data, but they were not necessarily involved in reflecting upon the collected data. Instead, their parents performed that task for them and conveyed the results. For example, Liam recorded his exercise and diet and relied on his father to compare it for him: \textit{``I give [the data] to my dad, so he compares it with the previous ones I give to him, and sometimes he conveys if, like, I'm doing good.''} In other cases, participants provided information to their parents, but they did not have access to the records. As Simone shared, \textit{``during breakfast, they ask me how many hours of sleep did you get... I feel they have written notes or something. I don't really ask them where they write those, and I don't know why they collect it.''} The participant was not able to extract insights from these records because she did not have access to them, though the process of answering the same question every morning helped her learn about how much sleep she needs each night to not feel fatigued. 

\subsubsection{Attending Doctors' Appointments Together}
The majority of participants attended doctors' appointments with their parents, which allowed them to practice communicating with doctors and learn about their conditions in a controlled environment where their parents could provide support if necessary. For example, as one participant shared, 

\begin{quote}
    \textit{``When it comes to answering the questions, my mom actually told me that I'm the one who's supposed to answer the questions because I'm the one with the condition... Maybe if the doctor wants to know some of the diets that I was supposed to take in the evening or the ones which my mom was preparing for me, that's what she can answer. But concerning my personal health, I'm the one who usually answers the questions.''} (Maya, diabetes) 
\end{quote}

As demonstrated here, parents will set expectations and extend opportunities for adolescents to take responsibility for their personal health. To answer questions at the doctor's office, Maya had to develop more awareness on a day-to-day basis about her health behaviors. Yet, she still included her mother in appointments in case she needed additional informational support, such as details about her diet at home. Though this participant was expected to try and answer questions at the doctor's office, some participants felt unprepared to attend appointments alone. As Hana expressed, \textit{``I don’t like it when I'm alone in there... I just had to answer all these questions. I'm like, oh, I don't know the answer to most of these... I had to ask my mom.''} Participants like Hana typically relied on parents to supply the information requested by providers, demonstrating the shared responsibility for illness management. Some participants even highlighted the benefit of including their parents in appointments from a health management perspective. As Simone shared, ``\textit{I feel like if I'm talking to my doctor, [my parents] know what I've been going through, or like what I've been experiencing, and they can also help me manage it better if they know what it's about.''} Through these appointments, parents not only supplemented with information and context about participants' experiences, but they gained insight into how to help them better manage their conditions. Liam and his father attended appointments together, and the participant relied on his father to interpret and demonstrate the doctor's directions regarding his diabetes: \textit{``it was not really easy for me to learn... the doctor showed my dad how to use [insulin], and my dad showed me at home.''} Likewise, Alisha shared that she felt \textit{``safe''} when her parents went to appointments with her, and that it is \textit{``more easy to comprehend sensitive information''} because her parents \textit{``know how to communicate''} with her.

\subsubsection{An Unclear Future}
It was not clear to participants what exactly independent health management looked like in practice, or how things would change for them. Some participants said that they had not discussed independence or healthcare transition with their parents at all (e.g., \textit{``that's a very stressful thought for me... we haven't talked about it''} -- Hana). Although parents and participants communicated openly about day-to-day health-related information, they seldom discussed long-term plans or expectations. When long-term plans were discussed, they were approached from a family perspective rather than the participant's perspective. One participant shared, \textit{``they tell me when maybe I live on my own, they really can't let me live on my own. So they'll be bringing one of my siblings to live with me''} (Sara). Participants expected to have more autonomy from their parents in the future, but the shared expectation seemed to be that the family would still be involved in illness management to some degree. In this case, responsibility would shift from Sara's parents to her siblings, as her parents worried about what would happen if she experienced an asthma attack while living alone. Another barrier to participants' transition to independence was an incomplete understanding of what health management entails in adulthood or what responsibilities they would have. Jon, for example, was not concerned about learning to manage his health for the future because \textit{``I always get this feeling... that [asthma] doesn't last forever},'' and \textit{``all I need to do is just be on my best behavior.''} Participants were used to following instructions from their parents or doctors, and they were uncertain about what health management looked like beyond that.

\section{Discussion}
The transition from pediatric to adult healthcare poses significant long-term implications for well-being. Adolescence is a critical stage in the healthcare transition process as patients gradually obtain more autonomy and self-management skills. Parent-adolescent collaboration is key to a successful transition, yet few studies within HCI and CSCW have examined collaborative practices between parents and adolescents specifically within the context of healthcare transition. In this section, we discuss how our findings could inform the design of health technologies that support adolescents' self-management skills and autonomy, while supporting parents' roles as co-managers and advisors. 

\subsection{Supporting Collaborative Illness Management in Adolescence}
In this section, we discuss technology design considerations around information-sharing and parent-adolescent collaboration to support adolescents' sense of agency and safety. 

\subsubsection{Designing for Disclosure and Control}
Our findings revealed that participants cared about ownership over their personal health data and were selective in sharing health-related information. In line with prior work~\cite{kaushansky_living_2017,van_der_velden_not_2013}, we found that participants sought to preserve a feeling of normalcy in day-to-day life by withholding health-related information from their peers. However, their desire to maintain privacy was sometimes compromised by their parents' unilateral decisions to involve their teachers and peers in illness management. Overriding adolescents' preferences for disclosure, even for the purpose of ensuring their safety, can be counterproductive to fostering independence by reducing adolescents' sense of control over their health. The degree to which adolescents believe that their own actions and behaviors determine their experiences (i.e., an internal locus of control~\cite{rotter_generalized_1966}) can have significant implications for their health behaviors~\cite{steptoe_locus_2001}. Though existing work examines adolescents' self-disclosure practices~\cite{holtz_t1dlookslikeme_2020,huh-yoo_help_2023,jin_understanding_2023}, little work has been done to explore adolescents' reactions to and understanding of involuntary disclosure-- that is, when adolescents' personal health information is disclosed to other parties without their consent. 
Our findings revealed that participants were uncomfortable when the result of disclosure was ongoing monitoring by their parents using their peers or teachers. That said, they were more accepting of disclosure when the intent was to educate others about their needs (e.g., an informational session conducted by parents to teach their peers about emergency response), even when they were not consulted beforehand. Buyuktur et al.~\cite{buyuktur_supporting_2018} observed similar conflicts of interest between individuals with Spinal Cord Injury and their caregivers, noting the impact of hierarchical relationships on collaboratively constructed independence. They assert that care networks are critical in shaping individuals' self-care practices, and by extension their independence and sense of agency. As adolescents age and the hierarchical nature of their relationship with parents begins to shift, their ability to control the privacy of their personal health information becomes even more important in managing relational boundaries with their families and peers~\cite{ebersole_taking_2016}. Ankrah et al.~\cite{a_ankrah_when_2022} defined relational boundaries as interpersonal expectations and agreements among people and highlighted the importance of control over disclosure as a way for cancer survivors to manage their self-presentation and relational boundaries. Healthcare technologies should support granular data sharing mechanisms that enable adolescents to maintain their boundaries. To support adolescents' sense of control, we emphasize the need to include them in the decision-making process and help them understand the benefits and risks of (non)disclosure, which can also prepare them for navigating the often difficult task of self-disclosure in adulthood (e.g., in the workplace~\cite{ganesh_work_2021}). Our results point to the need for nuanced discussion around consent and adolescents' right to control their personal health information, which can be complicated by parents' responsibility to ensure their children's physical safety. Below, we discuss the role that technology might play in balancing these needs.

\subsubsection{Balancing Independence and Safety}
We observed that participants often wanted to assume more responsibility for health management, yet they felt constricted by their parents' efforts to oversee their management activities as a means of ensuring their safety. While health technologies for younger children tend to emphasize continuous monitoring (e.g.,~\cite{hui_mammibelli_2012,wang_quantified_2017,westeyn_monitoring_2012}), complete and real-time access to health data can create friction in parent-adolescent relationships~\cite{yu_conflicts_2023,kaziunas_caring_2017}. As an alternative to remote monitoring, our findings point to opportunities for technology to \textbf{support the collaborative implementation of safeguards that reinforce adolescents' autonomy while ensuring their safety}. Barbarin et al.~\cite{barbarin_taking_2015} found that patients and their families safeguarded illness management activities by completing tasks together and creating feedback loops to support multi-level safety checks. Though this prior work demonstrates the value of safeguards in collaborative management, the focus was on adult patients. Adolescent-parent relationships have a different caregiving dynamic and safeguarding activities can be complicated by other tensions. For example, participants in our study viewed parental attempts to confirm or supervise their medication adherence less as a sign of caring and more so as a lack of trust in their management capabilities. In their study of seniors living in retirement communities, Caldeira et al.~\cite{caldeira_senior_2017} found that a daily check-in system that required seniors to press a button every morning allowed staff to monitor seniors' well-being in a minimally invasive manner. If seniors failed to complete the check-in within the designated time frame, then staff would first call them and then visit their apartment if they did not answer. A similar system that supports multi-level safety checks could be implemented for adolescents, reducing the need for active parental oversight and supporting adolescents' self-management practices. Given that nearly all participants had personal devices, a mobile health application could send automated reminders and provide a check-in feature to confirm medication adherence. Parents would only be notified in the event that adolescents did not complete the check-in, which could prompt them to text or call. Outsourcing the reminder and check-in processes to technology could also reduce adolescents' perceptions of parental ``nagging,'' which has been found to inhibit self-management~\cite{dashiff_parents_2011}. 

Outside of medication adherence, our findings suggest the need to design for responding to adolescents' medical emergencies, which necessitates discussion around privacy and data sharing. Cranor et al.~\cite{cranor_parents_2014} found that while parents and adolescents tend to agree that adolescents have some right to privacy, both parties believe that parents are justified in overriding that right to privacy in emergency situations. In a similar vein, Czeskis et al.~\cite{czeskis_parenting_2010} found that adolescents are more willing to share information (e.g., location) with their parents in emergency situations compared to non-emergency situations. Neither of these studies focused on health-related information, which can be particularly sensitive. Hong et al.~\cite{hong_using_2020} highlighted the need to develop data sharing models that balance adolescent patients' autonomy and safety, cautioning against the use of a strict open or closed data access policy. We found that although participants resisted close monitoring around their everyday activities, they understood the importance of keeping their parents informed in the event of emergencies.  Building on prior work, our results suggest that health tracking technologies can \textbf{support a data sharing model based on emergency versus non-emergency situations}. That is, tracking technologies can support adolescents in maintaining privacy and control over routine health information while providing mechanisms to release information in emergency contexts. For example, wearable devices can support the collection of routine health information and also serve as detection systems (e.g., for asthma~\cite{wu_ubi-asthma_2023}). In such a scenario, an adolescent's tracked data would not be visible to their parents without prior consent, but a sign of an asthma attack would immediately permit their parents to see vital signs. 

\subsection{Preparing Adolescents to Transition}
In this section, we discuss technological avenues to support adolescents' independent health management skills and prepare them to handle the responsibilities of adult healthcare. 

\subsubsection{Building Self-Efficacy}
Two of the most common barriers in the transition to adult care are adolescents' lack of self-management skills and knowledge deficits around their illness and the transition process~\cite{gray_barriers_2018}. Participants in our study experienced similar challenges, but they were motivated to learn about their conditions and better manage their health. Along with illness-related knowledge, motivation plays a critical role in children's transition to independent self-management~\cite{cha_transitioning_2022}. We observed that participants valued information from sources that felt more personal, such as YouTube and TikTok, which have been shown to facilitate social support and chronic illness management~\cite{zehrung2023self, huh2014health}. One participant preferred the use of a chatbot explicitly because it felt more personalized than other forms of information-seeking. Though only one participant mentioned the use of a chatbot, we see potential to \textbf{design and deploy chatbot assistants to provide personalized guidance and support} for adolescents with chronic illness. For example, AdolescentBot~\cite{rahman_adolescentbot_2021} was effective in answering sensitive health-related queries from adolescents, serving as the first line of support for adolescents seeking care. In our study, participants sometimes avoided asking their parents for help so as to not burden them. By interacting with a chatbot, adolescents can have the opportunity to improve their illness-related knowledge and attempt to solve challenges on their own before turning to their parents. If the conversational agent is unable to provide adequate support, then it could escalate the communication to parents or healthcare providers. For example, MD2Me~\cite{huang_preparing_2014} used text messages to reinforce illness management skills and provide a way for adolescents to report health concerns, which were then categorized as urgent or non-urgent and relayed accordingly to healthcare providers. For non-urgent health concerns, we suggest that the system requests adolescents' consent before forwarding their messages as a means of supporting their involvement in the decision-making process. In addition to providing informational and decision-making support, we envision that chatbots can provide emotional support to adolescents by leveraging large language models (LLMs). LLM-driven chatbots have the capacity to engage in free-form conversation with users and provide empathetic interactions in health contexts~\cite{jo_understanding_2023}, which can benefit adolescents as they deal with emotional challenges related to their condition (e.g., the loss of normalcy). Though we did not ask participants for their views on chatbots, we observed that they were interested in trying new technologies (e.g., downloading various mobile health applications and reading about smart monitoring devices). As such, we believe that they would find value in conversational agents, especially if they lack social support related to their illness. Future work could investigate how adolescents use chatbots to seek information and support, as well as how chatbots augment other forms of support received from family members and online communities. 

\subsubsection{Building Confidence in Patient-Provider Communication}
Another area where participants struggled was communicating with their providers, particularly without the assistance of their parents. The majority of youth with special healthcare needs do not receive transition planning support, which involves a healthcare provider working with youth to develop self-management skills and meeting with them alone during preventive visits~\cite{lebrun-harris_transition_2018}. Despite the importance of one-on-one meetings with providers, prior work has not addressed barriers to these meetings from adolescents' perspectives. We found that participants were hesitant to speak with providers by themselves because they relied upon their parents for informational and emotional support. While attending appointments with their parents can be critical in earlier stages of the transition process, we emphasize the importance of empowering adolescents to meet with providers alone. Technology can help\textbf{ prepare adolescents for appointments with their providers through collaborative tracking with their parents}. Prior work has demonstrated that patient-generated health data can guide patient-provider discussions during visits and support shared decision-making~\cite{mentis_crafting_2017,chung_more_2015}. Shared note-taking can engage adolescents in documenting and reflecting upon their symptoms with the support of their parents, as well as facilitate discussions around health~\cite{grimes_toward_2009} and motivate healthier behaviors~\cite{li_supporting_2020,lukoff_tablechat_2018}. We observed that shared note-taking and reflection provided participants with opportunities to build more complete health records and interpret their health behaviors. This practice can further serve as an opportunity for parents to teach adolescents self-management skills (e.g., interpreting recorded blood glucose levels) and the ability to identify trends in their symptoms. Engaging in collaborative reflection can enable adolescents to develop a deeper understanding of their medical history and be more prepared to answer providers' questions about their health. One participant even took a physical diary containing shared notes to appointments when her mother could not attend, suggesting that shared data can provide adolescents with the knowledge and confidence they need to meet with providers alone. We see possibilities for the design of applications that align with participants' existing practices of writing in shared diaries and posting sticky notes in shared spaces. For example, adolescents and parents could use their own mobile devices to create notes, which would then be shown on a tablet displayed in a central location in the home (e.g.,~\cite{pina_dreamcatcher_2020}). A tablet-based application could also support the co-creation of notes. The benefit of this design is that adolescents and parents would be able to add notes asynchronously from different locations, while still providing a shared display that facilitates collaboration and prompts discussion.

\subsection{Limitations}
We conducted interviews with adolescents diagnosed with different chronic illnesses. While this diversity allowed us to explore commonalities in illness management, it is important to acknowledge that there can be differences in management practices depending upon the specific condition. For instance, certain conditions such as asthma and diabetes are more prone to acute emergencies. As a result, the focus of medical management may shift more towards contingency planning. In addition, we did not collect demographic data on participants' area of residence. Healthcare systems vary by country and state, and participants' healthcare experiences might have been different according to their residence. Lastly, our study employed online recruiting methods and conducted interviews exclusively through Zoom. Online recruitment offers advantages in terms of including understudied and marginalized populations, especially youth~\cite{mcinroy_pitfalls_2016}. Furthermore, online recruitment has become more commonplace after the pandemic made traditional face-to-face recruitment and data collection methods challenging. That said, online recruitment can introduce a self-selection bias among participants, and there is an increasing concern regarding fraudulent participation in online research studies~\cite{salinas_are_2023,woolfall_identifying_2023}. To mitigate these issues, we implemented a simple verification process at the beginning of each interview, asking basic questions (e.g., year of birth) to compare against information collected from the screening questionnaire. If there was a discrepancy, we did not proceed with the interview. For future studies, we recommend implementing more rigorous verification procedures to ensure the integrity of the collected data, while still respecting participants' privacy. For example, researchers could use a secure third-party service to verify participants' identities before proceeding with the study.

\section{Conclusion}
Through interviews with 15 adolescents diagnosed with chronic illnesses, we found that adolescents assumed responsibility for learning about and performing mundane self-care tasks. Although they wanted to exert autonomy and maintain a sense of privacy, parents' efforts to ensure their safety sometimes led to tension (e.g., disclosing their illness to peers and teachers). Despite these tensions, participants and their parents engaged in the transition process together through shared activities such as attending doctors' appointments and record-keeping, which provided opportunities for adolescents to learn and practice self-management skills with parental support. Based on our findings, we presented considerations for designing technology to facilitate transition and promote parent-adolescent collaboration with respect to adolescents' needs and preferences.

\begin{acks}
We would like to thank our participants for sharing their experiences and insights, as well as reviewers for their detailed feedback. This work was supported by the National Science Foundation under Award IIS-2211923.

\end{acks}

\bibliographystyle{ACM-Reference-Format}
\bibliography{base}


\begin{thebibliography}{104}


\ifx \showCODEN    \undefined \def \showCODEN     #1{\unskip}     \fi
\ifx \showDOI      \undefined \def \showDOI       #1{#1}\fi
\ifx \showISBNx    \undefined \def \showISBNx     #1{\unskip}     \fi
\ifx \showISBNxiii \undefined \def \showISBNxiii  #1{\unskip}     \fi
\ifx \showISSN     \undefined \def \showISSN      #1{\unskip}     \fi
\ifx \showLCCN     \undefined \def \showLCCN      #1{\unskip}     \fi
\ifx \shownote     \undefined \def \shownote      #1{#1}          \fi
\ifx \showarticletitle \undefined \def \showarticletitle #1{#1}   \fi
\ifx \showURL      \undefined \def \showURL       {\relax}        \fi
\providecommand\bibfield[2]{#2}
\providecommand\bibinfo[2]{#2}
\providecommand\natexlab[1]{#1}
\providecommand\showeprint[2][]{arXiv:#2}

\bibitem[Abowd et~al\mbox{.}(2006)]%
        {abowd_2006_challenges}
\bibfield{author}{\bibinfo{person}{Gregory~D Abowd}, \bibinfo{person}{Gillian~R Hayes}, \bibinfo{person}{Julie~A Kientz}, \bibinfo{person}{Lena Mamykina}, {and} \bibinfo{person}{Elizabeth~D Mynatt}.} \bibinfo{year}{2006}\natexlab{}.
\newblock \showarticletitle{Challenges and opportunities for collaboration technologies for chronic care management}.
\newblock \bibinfo{journal}{\emph{The Human-Computer Interaction Consortium (HCIC 2006)}} (\bibinfo{year}{2006}).
\newblock


\bibitem[Amir et~al\mbox{.}(2015)]%
        {amir_care_2015}
\bibfield{author}{\bibinfo{person}{Ofra Amir}, \bibinfo{person}{Barbara~J. Grosz}, \bibinfo{person}{Krzysztof~Z. Gajos}, \bibinfo{person}{Sonja~M. Swenson}, {and} \bibinfo{person}{Lee~M. Sanders}.} \bibinfo{year}{2015}\natexlab{}.
\newblock \showarticletitle{From {Care} {Plans} to {Care} {Coordination}: {Opportunities} for {Computer} {Support} of {Teamwork} in {Complex} {Healthcare}}. In \bibinfo{booktitle}{\emph{Proceedings of the 33rd {Annual} {ACM} {Conference} on {Human} {Factors} in {Computing} {Systems}}} \emph{(\bibinfo{series}{{CHI} '15})}. \bibinfo{publisher}{Association for Computing Machinery}, \bibinfo{address}{New York, NY, USA}, \bibinfo{pages}{1419--1428}.
\newblock
\showISBNx{978-1-4503-3145-6}
\urldef\tempurl%
\url{https://doi.org/10.1145/2702123.2702320}
\showDOI{\tempurl}


\bibitem[Ancker et~al\mbox{.}(2015)]%
        {ancker_you_2015}
\bibfield{author}{\bibinfo{person}{Jessica~S. Ancker}, \bibinfo{person}{Holly~O. Witteman}, \bibinfo{person}{Baria Hafeez}, \bibinfo{person}{Thierry Provencher}, \bibinfo{person}{Mary Van~de Graaf}, {and} \bibinfo{person}{Esther Wei}.} \bibinfo{year}{2015}\natexlab{}.
\newblock \showarticletitle{“{You} {Get} {Reminded} {You}’re a {Sick} {Person}”: {Personal} {Data} {Tracking} and {Patients} {With} {Multiple} {Chronic} {Conditions}}.
\newblock \bibinfo{journal}{\emph{Journal of Medical Internet Research}} \bibinfo{volume}{17}, \bibinfo{number}{8} (\bibinfo{date}{Aug.} \bibinfo{year}{2015}), \bibinfo{pages}{e4209}.
\newblock
\urldef\tempurl%
\url{https://doi.org/10.2196/jmir.4209}
\showDOI{\tempurl}


\bibitem[Ankrah et~al\mbox{.}(2022a)]%
        {a_ankrah_when_2022}
\bibfield{author}{\bibinfo{person}{Elizabeth~A. Ankrah}, \bibinfo{person}{Arpita Bhattacharya}, \bibinfo{person}{Lissamarie Donjuan}, \bibinfo{person}{Franceli~L. Cibrian}, \bibinfo{person}{Lilibeth Torno}, \bibinfo{person}{Anamara Ritt~Olson}, \bibinfo{person}{Joel Milam}, {and} \bibinfo{person}{Gillian Hayes}.} \bibinfo{year}{2022}\natexlab{a}.
\newblock \showarticletitle{When {Worlds} {Collide}: {Boundary} {Management} of {Adolescent} and {Young} {Adult} {Childhood} {Cancer} {Survivors} and {Caregivers}}. In \bibinfo{booktitle}{\emph{{CHI} {Conference} on {Human} {Factors} in {Computing} {Systems}}}. \bibinfo{publisher}{ACM}, \bibinfo{address}{New Orleans LA USA}, \bibinfo{pages}{1--16}.
\newblock
\showISBNx{978-1-4503-9157-3}
\urldef\tempurl%
\url{https://doi.org/10.1145/3491102.3517544}
\showDOI{\tempurl}


\bibitem[Ankrah et~al\mbox{.}(2022b)]%
        {ankrah_me_2022}
\bibfield{author}{\bibinfo{person}{Elizabeth~A. Ankrah}, \bibinfo{person}{Franceli~L. Cibrian}, \bibinfo{person}{Lucas~M. Silva}, \bibinfo{person}{Arya Tavakoulnia}, \bibinfo{person}{Jesus~A. Beltran}, \bibinfo{person}{Sabrina~E.B. Schuck}, \bibinfo{person}{Kimberley~D. Lakes}, {and} \bibinfo{person}{Gillian~R. Hayes}.} \bibinfo{year}{2022}\natexlab{b}.
\newblock \showarticletitle{Me, {My} {Health}, and {My} {Watch}: {How} {Children} with {ADHD} {Understand} {Smartwatch} {Health} {Data}}.
\newblock \bibinfo{journal}{\emph{ACM Transactions on Computer-Human Interaction}} (\bibinfo{date}{Dec.} \bibinfo{year}{2022}), \bibinfo{pages}{3577008}.
\newblock
\showISSN{1073-0516, 1557-7325}
\urldef\tempurl%
\url{https://doi.org/10.1145/3577008}
\showDOI{\tempurl}


\bibitem[Ayobi et~al\mbox{.}(2020)]%
        {ayobi_trackly_2020}
\bibfield{author}{\bibinfo{person}{Amid Ayobi}, \bibinfo{person}{Paul Marshall}, {and} \bibinfo{person}{Anna~L. Cox}.} \bibinfo{year}{2020}\natexlab{}.
\newblock \showarticletitle{Trackly: {A} {Customisable} and {Pictorial} {Self}-{Tracking} {App} to {Support} {Agency} in {Multiple} {Sclerosis} {Self}-{Care}}. In \bibinfo{booktitle}{\emph{Proceedings of the 2020 {CHI} {Conference} on {Human} {Factors} in {Computing} {Systems}}} \emph{(\bibinfo{series}{{CHI} '20})}. \bibinfo{publisher}{Association for Computing Machinery}, \bibinfo{address}{New York, NY, USA}, \bibinfo{pages}{1--15}.
\newblock
\showISBNx{978-1-4503-6708-0}
\urldef\tempurl%
\url{https://doi.org/10.1145/3313831.3376809}
\showDOI{\tempurl}


\bibitem[Ayobi et~al\mbox{.}(2018)]%
        {ayobi_flexible_2018}
\bibfield{author}{\bibinfo{person}{Amid Ayobi}, \bibinfo{person}{Tobias Sonne}, \bibinfo{person}{Paul Marshall}, {and} \bibinfo{person}{Anna~L. Cox}.} \bibinfo{year}{2018}\natexlab{}.
\newblock \showarticletitle{Flexible and {Mindful} {Self}-{Tracking}: {Design} {Implications} from {Paper} {Bullet} {Journals}}. In \bibinfo{booktitle}{\emph{Proceedings of the 2018 {CHI} {Conference} on {Human} {Factors} in {Computing} {Systems}}} \emph{(\bibinfo{series}{{CHI} '18})}. \bibinfo{publisher}{Association for Computing Machinery}, \bibinfo{address}{New York, NY, USA}, \bibinfo{pages}{1--14}.
\newblock
\showISBNx{978-1-4503-5620-6}
\urldef\tempurl%
\url{https://doi.org/10.1145/3173574.3173602}
\showDOI{\tempurl}


\bibitem[Babler and Strickland(2015)]%
        {babler_moving_2015}
\bibfield{author}{\bibinfo{person}{Elizabeth Babler} {and} \bibinfo{person}{Carolyn~June Strickland}.} \bibinfo{year}{2015}\natexlab{}.
\newblock \showarticletitle{Moving the {Journey} {Towards} {Independence}: {Adolescents} {Transitioning} to {Successful} {Diabetes} {Self}-{Management}}.
\newblock \bibinfo{journal}{\emph{Journal of Pediatric Nursing}} \bibinfo{volume}{30}, \bibinfo{number}{5} (\bibinfo{date}{Sept.} \bibinfo{year}{2015}), \bibinfo{pages}{648--660}.
\newblock
\showISSN{0882-5963}
\urldef\tempurl%
\url{https://doi.org/10.1016/j.pedn.2015.06.005}
\showDOI{\tempurl}


\bibitem[Barbarin et~al\mbox{.}(2015)]%
        {barbarin_taking_2015}
\bibfield{author}{\bibinfo{person}{Andrea Barbarin}, \bibinfo{person}{Tiffany~C. Veinot}, {and} \bibinfo{person}{Predrag Klasnja}.} \bibinfo{year}{2015}\natexlab{}.
\newblock \showarticletitle{Taking our {Time}: {Chronic} {Illness} and {Time}-{Based} {Objects} in {Families}}. In \bibinfo{booktitle}{\emph{Proceedings of the 18th {ACM} {Conference} on {Computer} {Supported} {Cooperative} {Work} \& {Social} {Computing}}} \emph{(\bibinfo{series}{{CSCW} '15})}. \bibinfo{publisher}{Association for Computing Machinery}, \bibinfo{address}{New York, NY, USA}, \bibinfo{pages}{288--301}.
\newblock
\showISBNx{978-1-4503-2922-4}
\urldef\tempurl%
\url{https://doi.org/10.1145/2675133.2675200}
\showDOI{\tempurl}


\bibitem[Barlow et~al\mbox{.}(2002)]%
        {barlow_self-management_2002}
\bibfield{author}{\bibinfo{person}{Julie Barlow}, \bibinfo{person}{Chris Wright}, \bibinfo{person}{Janice Sheasby}, \bibinfo{person}{Andy Turner}, {and} \bibinfo{person}{Jenny Hainsworth}.} \bibinfo{year}{2002}\natexlab{}.
\newblock \showarticletitle{Self-management approaches for people with chronic conditions: a review}.
\newblock \bibinfo{journal}{\emph{Patient Education and Counseling}} \bibinfo{volume}{48}, \bibinfo{number}{2} (\bibinfo{date}{Oct.} \bibinfo{year}{2002}), \bibinfo{pages}{177--187}.
\newblock
\showISSN{0738-3991}
\urldef\tempurl%
\url{https://doi.org/10.1016/S0738-3991(02)00032-0}
\showDOI{\tempurl}


\bibitem[Barlow and Ellard(2006)]%
        {barlow_psychosocial_2006}
\bibfield{author}{\bibinfo{person}{J.~H. Barlow} {and} \bibinfo{person}{D.~R. Ellard}.} \bibinfo{year}{2006}\natexlab{}.
\newblock \showarticletitle{The psychosocial well-being of children with chronic disease, their parents and siblings: an overview of the research evidence base}.
\newblock \bibinfo{journal}{\emph{Child: Care, Health and Development}} \bibinfo{volume}{32}, \bibinfo{number}{1} (\bibinfo{year}{2006}), \bibinfo{pages}{19--31}.
\newblock
\showISSN{1365-2214}
\urldef\tempurl%
\url{https://doi.org/10.1111/j.1365-2214.2006.00591.x}
\showDOI{\tempurl}


\bibitem[Beacham and Deatrick(2015)]%
        {beacham_children_2015}
\bibfield{author}{\bibinfo{person}{Barbara~L. Beacham} {and} \bibinfo{person}{Janet~A. Deatrick}.} \bibinfo{year}{2015}\natexlab{}.
\newblock \showarticletitle{Children {With} {Chronic} {Conditions}: {Perspectives} on {Condition} {Management}}.
\newblock \bibinfo{journal}{\emph{Journal of Pediatric Nursing}} \bibinfo{volume}{30}, \bibinfo{number}{1} (\bibinfo{date}{Jan.} \bibinfo{year}{2015}), \bibinfo{pages}{25--35}.
\newblock
\showISSN{0882-5963}
\urldef\tempurl%
\url{https://doi.org/10.1016/j.pedn.2014.10.011}
\showDOI{\tempurl}


\bibitem[Bhat et~al\mbox{.}(2023)]%
        {bhat_we_2023}
\bibfield{author}{\bibinfo{person}{Karthik~S. Bhat}, \bibinfo{person}{Amanda~K. Hall}, \bibinfo{person}{Tiffany Kuo}, {and} \bibinfo{person}{Neha Kumar}.} \bibinfo{year}{2023}\natexlab{}.
\newblock \showarticletitle{``{We} are half-doctors'': {Family} {Caregivers} as {Boundary} {Actors} in {Chronic} {Disease} {Management}}.
\newblock \bibinfo{journal}{\emph{Proceedings of the ACM on Human-Computer Interaction}} \bibinfo{volume}{7}, \bibinfo{number}{CSCW1} (\bibinfo{date}{April} \bibinfo{year}{2023}), \bibinfo{pages}{111:1--111:29}.
\newblock
\urldef\tempurl%
\url{https://doi.org/10.1145/3579545}
\showDOI{\tempurl}


\bibitem[Branje(2018)]%
        {branje_development_2018}
\bibfield{author}{\bibinfo{person}{Susan Branje}.} \bibinfo{year}{2018}\natexlab{}.
\newblock \showarticletitle{Development of {Parent}–{Adolescent} {Relationships}: {Conflict} {Interactions} as a {Mechanism} of {Change}}.
\newblock \bibinfo{journal}{\emph{Child Development Perspectives}} \bibinfo{volume}{12}, \bibinfo{number}{3} (\bibinfo{year}{2018}), \bibinfo{pages}{171--176}.
\newblock
\showISSN{1750-8606}
\urldef\tempurl%
\url{https://doi.org/10.1111/cdep.12278}
\showDOI{\tempurl}


\bibitem[Braun and Clarke(2012)]%
        {braun_thematic_2012}
\bibfield{author}{\bibinfo{person}{Virginia Braun} {and} \bibinfo{person}{Victoria Clarke}.} \bibinfo{year}{2012}\natexlab{}.
\newblock \showarticletitle{Thematic analysis}.
\newblock In \bibinfo{booktitle}{\emph{{APA} handbook of research methods in psychology, {Vol} 2: {Research} designs: {Quantitative}, qualitative, neuropsychological, and biological}}. \bibinfo{publisher}{American Psychological Association}, \bibinfo{address}{Washington, DC, US}, \bibinfo{pages}{57--71}.
\newblock
\showISBNx{978-1-4338-1005-3}
\urldef\tempurl%
\url{https://doi.org/10.1037/13620-004}
\showDOI{\tempurl}


\bibitem[Brousseau et~al\mbox{.}(2010)]%
        {brousseau_acute_2010}
\bibfield{author}{\bibinfo{person}{David~C. Brousseau}, \bibinfo{person}{Pamela~L. Owens}, \bibinfo{person}{Andrew~L. Mosso}, \bibinfo{person}{Julie~A. Panepinto}, {and} \bibinfo{person}{Claudia~A. Steiner}.} \bibinfo{year}{2010}\natexlab{}.
\newblock \showarticletitle{Acute {Care} {Utilization} and {Rehospitalizations} for {Sickle} {Cell} {Disease}}.
\newblock \bibinfo{journal}{\emph{JAMA}} \bibinfo{volume}{303}, \bibinfo{number}{13} (\bibinfo{date}{April} \bibinfo{year}{2010}), \bibinfo{pages}{1288--1294}.
\newblock
\showISSN{0098-7484}
\urldef\tempurl%
\url{https://doi.org/10.1001/jama.2010.378}
\showDOI{\tempurl}


\bibitem[Brown and Bussell(2011)]%
        {brown_medication_2011}
\bibfield{author}{\bibinfo{person}{Marie~T. Brown} {and} \bibinfo{person}{Jennifer~K. Bussell}.} \bibinfo{year}{2011}\natexlab{}.
\newblock \showarticletitle{Medication {Adherence}: {WHO} {Cares}?}
\newblock \bibinfo{journal}{\emph{Mayo Clinic Proceedings}} \bibinfo{volume}{86}, \bibinfo{number}{4} (\bibinfo{date}{April} \bibinfo{year}{2011}), \bibinfo{pages}{304--314}.
\newblock
\showISSN{0025-6196}
\urldef\tempurl%
\url{https://doi.org/10.4065/mcp.2010.0575}
\showDOI{\tempurl}


\bibitem[Buyuktur et~al\mbox{.}(2018)]%
        {buyuktur_supporting_2018}
\bibfield{author}{\bibinfo{person}{Ayse~G. Buyuktur}, \bibinfo{person}{Pei-Yao Hung}, \bibinfo{person}{Mark~W. Newman}, {and} \bibinfo{person}{Mark~S. Ackerman}.} \bibinfo{year}{2018}\natexlab{}.
\newblock \showarticletitle{Supporting {Collaboratively} {Constructed} {Independence}: {A} {Study} of {Spinal} {Cord} {Injury}}.
\newblock \bibinfo{journal}{\emph{Proceedings of the ACM on Human-Computer Interaction}} \bibinfo{volume}{2}, \bibinfo{number}{CSCW} (\bibinfo{date}{Nov.} \bibinfo{year}{2018}), \bibinfo{pages}{26:1--26:25}.
\newblock
\urldef\tempurl%
\url{https://doi.org/10.1145/3274295}
\showDOI{\tempurl}


\bibitem[Caldeira et~al\mbox{.}(2017)]%
        {caldeira_senior_2017}
\bibfield{author}{\bibinfo{person}{Clara Caldeira}, \bibinfo{person}{Matthew Bietz}, \bibinfo{person}{Marisol Vidauri}, {and} \bibinfo{person}{Yunan Chen}.} \bibinfo{year}{2017}\natexlab{}.
\newblock \showarticletitle{Senior {Care} for {Aging} in {Place}: {Balancing} {Assistance} and {Independence}}. In \bibinfo{booktitle}{\emph{Proceedings of the 2017 {ACM} {Conference} on {Computer} {Supported} {Cooperative} {Work} and {Social} {Computing}}} \emph{(\bibinfo{series}{{CSCW} '17})}. \bibinfo{publisher}{Association for Computing Machinery}, \bibinfo{address}{New York, NY, USA}, \bibinfo{pages}{1605--1617}.
\newblock
\showISBNx{978-1-4503-4335-0}
\urldef\tempurl%
\url{https://doi.org/10.1145/2998181.2998206}
\showDOI{\tempurl}


\bibitem[Cha et~al\mbox{.}(2022)]%
        {cha_transitioning_2022}
\bibfield{author}{\bibinfo{person}{Yoon~Jeong Cha}, \bibinfo{person}{Arpita Saxena}, \bibinfo{person}{Alice Wou}, \bibinfo{person}{Joyce Lee}, \bibinfo{person}{Mark~W Newman}, {and} \bibinfo{person}{Sun~Young Park}.} \bibinfo{year}{2022}\natexlab{}.
\newblock \showarticletitle{Transitioning {Toward} {Independence}: {Enhancing} {Collaborative} {Self}-{Management} of {Children} with {Type} 1 {Diabetes}}. In \bibinfo{booktitle}{\emph{{CHI} {Conference} on {Human} {Factors} in {Computing} {Systems}}} \emph{(\bibinfo{series}{{CHI} '22})}. \bibinfo{publisher}{Association for Computing Machinery}, \bibinfo{address}{New York, NY, USA}, \bibinfo{pages}{1--17}.
\newblock
\showISBNx{978-1-4503-9157-3}
\urldef\tempurl%
\url{https://doi.org/10.1145/3491102.3502055}
\showDOI{\tempurl}


\bibitem[Chauke et~al\mbox{.}(2022)]%
        {chauke_factors_2022}
\bibfield{author}{\bibinfo{person}{Gloria~Dunisani Chauke}, \bibinfo{person}{Olivia Nakwafila}, \bibinfo{person}{Buyisile Chibi}, \bibinfo{person}{Benn Sartorius}, {and} \bibinfo{person}{Tivani Mashamba-Thompson}.} \bibinfo{year}{2022}\natexlab{}.
\newblock \showarticletitle{Factors influencing poor medication adherence amongst patients with chronic disease in low-and-middle-income countries: {A} systematic scoping review}.
\newblock \bibinfo{journal}{\emph{Heliyon}} \bibinfo{volume}{8}, \bibinfo{number}{6} (\bibinfo{date}{June} \bibinfo{year}{2022}), \bibinfo{pages}{e09716}.
\newblock
\showISSN{2405-8440}
\urldef\tempurl%
\url{https://doi.org/10.1016/j.heliyon.2022.e09716}
\showDOI{\tempurl}


\bibitem[Chopra et~al\mbox{.}(2021)]%
        {chopra_living_2021}
\bibfield{author}{\bibinfo{person}{Shaan Chopra}, \bibinfo{person}{Rachael Zehrung}, \bibinfo{person}{Tamil~Arasu Shanmugam}, {and} \bibinfo{person}{Eun~Kyoung Choe}.} \bibinfo{year}{2021}\natexlab{}.
\newblock \showarticletitle{Living with {Uncertainty} and {Stigma}: {Self}-{Experimentation} and {Support}-{Seeking} around {Polycystic} {Ovary} {Syndrome}}. In \bibinfo{booktitle}{\emph{Proceedings of the 2021 {CHI} {Conference} on {Human} {Factors} in {Computing} {Systems}}} \emph{(\bibinfo{series}{{CHI} '21})}. \bibinfo{publisher}{Association for Computing Machinery}, \bibinfo{address}{New York, NY, USA}, \bibinfo{pages}{1--18}.
\newblock
\showISBNx{978-1-4503-8096-6}
\urldef\tempurl%
\url{https://doi.org/10.1145/3411764.3445706}
\showDOI{\tempurl}


\bibitem[Chung et~al\mbox{.}(2015)]%
        {chung_more_2015}
\bibfield{author}{\bibinfo{person}{Chia-Fang Chung}, \bibinfo{person}{Jonathan Cook}, \bibinfo{person}{Elizabeth Bales}, \bibinfo{person}{Jasmine Zia}, {and} \bibinfo{person}{Sean~A. Munson}.} \bibinfo{year}{2015}\natexlab{}.
\newblock \showarticletitle{More {Than} {Telemonitoring}: {Health} {Provider} {Use} and {Nonuse} of {Life}-{Log} {Data} in {Irritable} {Bowel} {Syndrome} and {Weight} {Management}}.
\newblock \bibinfo{journal}{\emph{Journal of Medical Internet Research}} \bibinfo{volume}{17}, \bibinfo{number}{8} (\bibinfo{date}{Aug.} \bibinfo{year}{2015}), \bibinfo{pages}{e4364}.
\newblock
\urldef\tempurl%
\url{https://doi.org/10.2196/jmir.4364}
\showDOI{\tempurl}


\bibitem[Chung et~al\mbox{.}(2016)]%
        {chung_boundary_2016}
\bibfield{author}{\bibinfo{person}{Chia-Fang Chung}, \bibinfo{person}{Kristin Dew}, \bibinfo{person}{Allison Cole}, \bibinfo{person}{Jasmine Zia}, \bibinfo{person}{James Fogarty}, \bibinfo{person}{Julie~A. Kientz}, {and} \bibinfo{person}{Sean~A. Munson}.} \bibinfo{year}{2016}\natexlab{}.
\newblock \showarticletitle{Boundary {Negotiating} {Artifacts} in {Personal} {Informatics}: {Patient}-{Provider} {Collaboration} with {Patient}-{Generated} {Data}}. In \bibinfo{booktitle}{\emph{Proceedings of the 19th {ACM} {Conference} on {Computer}-{Supported} {Cooperative} {Work} \& {Social} {Computing}}} \emph{(\bibinfo{series}{{CSCW} '16})}. \bibinfo{publisher}{Association for Computing Machinery}, \bibinfo{address}{New York, NY, USA}, \bibinfo{pages}{770--786}.
\newblock
\showISBNx{978-1-4503-3592-8}
\urldef\tempurl%
\url{https://doi.org/10.1145/2818048.2819926}
\showDOI{\tempurl}


\bibitem[Colineau and Paris(2011)]%
        {colineau_motivating_2011}
\bibfield{author}{\bibinfo{person}{Nathalie Colineau} {and} \bibinfo{person}{Cécile Paris}.} \bibinfo{year}{2011}\natexlab{}.
\newblock \showarticletitle{Motivating reflection about health within the family: the use of goal setting and tailored feedback}.
\newblock \bibinfo{journal}{\emph{User Modeling and User-Adapted Interaction}} \bibinfo{volume}{21}, \bibinfo{number}{4} (\bibinfo{date}{Oct.} \bibinfo{year}{2011}), \bibinfo{pages}{341--376}.
\newblock
\showISSN{1573-1391}
\urldef\tempurl%
\url{https://doi.org/10.1007/s11257-010-9089-x}
\showDOI{\tempurl}


\bibitem[Cranor et~al\mbox{.}(2014)]%
        {cranor_parents_2014}
\bibfield{author}{\bibinfo{person}{Lorrie~Faith Cranor}, \bibinfo{person}{Adam~L. Durity}, \bibinfo{person}{Abigail Marsh}, {and} \bibinfo{person}{Blase Ur}.} \bibinfo{year}{2014}\natexlab{}.
\newblock \showarticletitle{{Parents{\textquoteright}} and {Teens{\textquoteright}} Perspectives on Privacy In a {Technology-Filled} World}. In \bibinfo{booktitle}{\emph{10th Symposium On Usable Privacy and Security (SOUPS 2014)}}. \bibinfo{publisher}{USENIX Association}, \bibinfo{address}{Menlo Park, CA}, \bibinfo{pages}{19--35}.
\newblock
\showISBNx{978-1-931971-13-3}
\urldef\tempurl%
\url{https://www.usenix.org/conference/soups2014/proceedings/presentation/cranor}
\showURL{%
\tempurl}


\bibitem[Czeskis et~al\mbox{.}(2010)]%
        {czeskis_parenting_2010}
\bibfield{author}{\bibinfo{person}{Alexei Czeskis}, \bibinfo{person}{Ivayla Dermendjieva}, \bibinfo{person}{Hussein Yapit}, \bibinfo{person}{Alan Borning}, \bibinfo{person}{Batya Friedman}, \bibinfo{person}{Brian Gill}, {and} \bibinfo{person}{Tadayoshi Kohno}.} \bibinfo{year}{2010}\natexlab{}.
\newblock \showarticletitle{Parenting from the pocket: value tensions and technical directions for secure and private parent-teen mobile safety}. In \bibinfo{booktitle}{\emph{Proceedings of the {Sixth} {Symposium} on {Usable} {Privacy} and {Security}}} \emph{(\bibinfo{series}{{SOUPS} '10})}. \bibinfo{publisher}{Association for Computing Machinery}, \bibinfo{address}{New York, NY, USA}, \bibinfo{pages}{1--15}.
\newblock
\showISBNx{978-1-4503-0264-7}
\urldef\tempurl%
\url{https://doi.org/10.1145/1837110.1837130}
\showDOI{\tempurl}


\bibitem[Dashiff et~al\mbox{.}(2011)]%
        {dashiff_parents_2011}
\bibfield{author}{\bibinfo{person}{Carol Dashiff}, \bibinfo{person}{Bettina~H. Riley}, \bibinfo{person}{Hussein Abdullatif}, {and} \bibinfo{person}{Elaine Moreland}.} \bibinfo{year}{2011}\natexlab{}.
\newblock \showarticletitle{Parents' experiences supporting self-management of middle adolescents with type 1 diabetes mellitus}.
\newblock \bibinfo{journal}{\emph{Pediatric Nursing}} \bibinfo{volume}{37}, \bibinfo{number}{6} (\bibinfo{date}{Dec.} \bibinfo{year}{2011}), \bibinfo{pages}{304--310}.
\newblock
\showISSN{0097-9805}


\bibitem[Davis et~al\mbox{.}(2018)]%
        {davis_kiss_2018}
\bibfield{author}{\bibinfo{person}{SR Davis}, \bibinfo{person}{D Peters}, \bibinfo{person}{RA Calvo}, \bibinfo{person}{SM Sawyer}, \bibinfo{person}{JM Foster}, {and} \bibinfo{person}{L Smith}.} \bibinfo{year}{2018}\natexlab{}.
\newblock \showarticletitle{“{Kiss} {myAsthma}”: {Using} a participatory design approach to develop a self-management app with young people with asthma}.
\newblock \bibinfo{journal}{\emph{Journal of Asthma}} \bibinfo{volume}{55}, \bibinfo{number}{9} (\bibinfo{date}{Sept.} \bibinfo{year}{2018}), \bibinfo{pages}{1018--1027}.
\newblock
\showISSN{0277-0903}
\urldef\tempurl%
\url{https://doi.org/10.1080/02770903.2017.1388391}
\showDOI{\tempurl}


\bibitem[Dean et~al\mbox{.}(2010)]%
        {dean_systematic_2010}
\bibfield{author}{\bibinfo{person}{Angela~J. Dean}, \bibinfo{person}{Julie Walters}, {and} \bibinfo{person}{Anthony Hall}.} \bibinfo{year}{2010}\natexlab{}.
\newblock \showarticletitle{A systematic review of interventions to enhance medication adherence in children and adolescents with chronic illness}.
\newblock \bibinfo{journal}{\emph{Archives of Disease in Childhood}} \bibinfo{volume}{95}, \bibinfo{number}{9} (\bibinfo{date}{Sept.} \bibinfo{year}{2010}), \bibinfo{pages}{717--723}.
\newblock
\showISSN{0003-9888, 1468-2044}
\urldef\tempurl%
\url{https://doi.org/10.1136/adc.2009.175125}
\showDOI{\tempurl}


\bibitem[Desai et~al\mbox{.}(2019)]%
        {desai_personal_2019}
\bibfield{author}{\bibinfo{person}{Pooja~M. Desai}, \bibinfo{person}{Elliot~G. Mitchell}, \bibinfo{person}{Maria~L. Hwang}, \bibinfo{person}{Matthew~E. Levine}, \bibinfo{person}{David~J. Albers}, {and} \bibinfo{person}{Lena Mamykina}.} \bibinfo{year}{2019}\natexlab{}.
\newblock \showarticletitle{Personal {Health} {Oracle}: {Explorations} of {Personalized} {Predictions} in {Diabetes} {Self}-{Management}}. In \bibinfo{booktitle}{\emph{Proceedings of the 2019 {CHI} {Conference} on {Human} {Factors} in {Computing} {Systems}}} \emph{(\bibinfo{series}{{CHI} '19})}. \bibinfo{publisher}{Association for Computing Machinery}, \bibinfo{address}{New York, NY, USA}, \bibinfo{pages}{1--13}.
\newblock
\showISBNx{978-1-4503-5970-2}
\urldef\tempurl%
\url{https://doi.org/10.1145/3290605.3300600}
\showDOI{\tempurl}


\bibitem[Ebersole and Hernandez(2016)]%
        {ebersole_taking_2016}
\bibfield{author}{\bibinfo{person}{Diana~S. Ebersole} {and} \bibinfo{person}{Rachael~A. Hernandez}.} \bibinfo{year}{2016}\natexlab{}.
\newblock \showarticletitle{“{Taking} {Good} {Care} of {Our} {Health}”: {Parent}-{Adolescent} {Perceptions} of {Boundary} {Management} {About} {Health} {Information}}.
\newblock \bibinfo{journal}{\emph{Communication Quarterly}} \bibinfo{volume}{64}, \bibinfo{number}{5} (\bibinfo{date}{Oct.} \bibinfo{year}{2016}), \bibinfo{pages}{573--595}.
\newblock
\showISSN{0146-3373}
\urldef\tempurl%
\url{https://doi.org/10.1080/01463373.2016.1176939}
\showDOI{\tempurl}


\bibitem[Epstein et~al\mbox{.}(2020)]%
        {epstein_mapping_2020}
\bibfield{author}{\bibinfo{person}{Daniel~A. Epstein}, \bibinfo{person}{Clara Caldeira}, \bibinfo{person}{Mayara~Costa Figueiredo}, \bibinfo{person}{Xi Lu}, \bibinfo{person}{Lucas~M. Silva}, \bibinfo{person}{Lucretia Williams}, \bibinfo{person}{Jong~Ho Lee}, \bibinfo{person}{Qingyang Li}, \bibinfo{person}{Simran Ahuja}, \bibinfo{person}{Qiuer Chen}, \bibinfo{person}{Payam Dowlatyari}, \bibinfo{person}{Craig Hilby}, \bibinfo{person}{Sazeda Sultana}, \bibinfo{person}{Elizabeth~V. Eikey}, {and} \bibinfo{person}{Yunan Chen}.} \bibinfo{year}{2020}\natexlab{}.
\newblock \showarticletitle{Mapping and {Taking} {Stock} of the {Personal} {Informatics} {Literature}}.
\newblock \bibinfo{journal}{\emph{Proceedings of the ACM on Interactive, Mobile, Wearable and Ubiquitous Technologies}} \bibinfo{volume}{4}, \bibinfo{number}{4} (\bibinfo{date}{Dec.} \bibinfo{year}{2020}), \bibinfo{pages}{126:1--126:38}.
\newblock
\urldef\tempurl%
\url{https://doi.org/10.1145/3432231}
\showDOI{\tempurl}


\bibitem[for Chronic Disease~Prevention and Promotion(2022)]%
        {cdc_about_2022}
\bibfield{author}{\bibinfo{person}{National~Center for Chronic Disease~Prevention} {and} \bibinfo{person}{Health Promotion}.} \bibinfo{year}{2022}\natexlab{}.
\newblock \bibinfo{title}{About {Chronic} {Diseases} {\textbar} {CDC}}.
\newblock
\newblock
\urldef\tempurl%
\url{https://www.cdc.gov/chronicdisease/about/index.htm}
\showURL{%
\tempurl}


\bibitem[for Disease~Control and Prevention(2022)]%
        {cdc_managing_2022}
\bibfield{author}{\bibinfo{person}{Centers for Disease~Control} {and} \bibinfo{person}{Prevention}.} \bibinfo{year}{2022}\natexlab{}.
\newblock \bibinfo{title}{Managing {Chronic} {Health} {Conditions} in {Schools} {\textbar} {Healthy} {Schools} {\textbar} {CDC}}.
\newblock
\newblock
\urldef\tempurl%
\url{https://www.cdc.gov/healthyschools/chronicconditions.htm}
\showURL{%
\tempurl}


\bibitem[Ganesh and Lazar(2021)]%
        {ganesh_work_2021}
\bibfield{author}{\bibinfo{person}{Kausalya Ganesh} {and} \bibinfo{person}{Amanda Lazar}.} \bibinfo{year}{2021}\natexlab{}.
\newblock \showarticletitle{The {Work} of {Workplace} {Disclosure}: {Invisible} {Chronic} {Conditions} and {Opportunities} for {Design}}.
\newblock \bibinfo{journal}{\emph{Proceedings of the ACM on Human-Computer Interaction}} \bibinfo{volume}{5}, \bibinfo{number}{CSCW1} (\bibinfo{date}{April} \bibinfo{year}{2021}), \bibinfo{pages}{73:1--73:26}.
\newblock
\urldef\tempurl%
\url{https://doi.org/10.1145/3449147}
\showDOI{\tempurl}


\bibitem[Garvey et~al\mbox{.}(2012)]%
        {garvey_health_2012}
\bibfield{author}{\bibinfo{person}{Katharine~C. Garvey}, \bibinfo{person}{Howard~A. Wolpert}, \bibinfo{person}{Erinn~T. Rhodes}, \bibinfo{person}{Lori~M. Laffel}, \bibinfo{person}{Ken Kleinman}, \bibinfo{person}{Margaret~G. Beste}, \bibinfo{person}{Joseph~I. Wolfsdorf}, {and} \bibinfo{person}{Jonathan~A. Finkelstein}.} \bibinfo{year}{2012}\natexlab{}.
\newblock \showarticletitle{Health {Care} {Transition} in {Patients} {With} {Type} 1 {Diabetes}: {Young} adult experiences and relationship to glycemic control}.
\newblock \bibinfo{journal}{\emph{Diabetes Care}} \bibinfo{volume}{35}, \bibinfo{number}{8} (\bibinfo{date}{July} \bibinfo{year}{2012}), \bibinfo{pages}{1716--1722}.
\newblock
\showISSN{0149-5992}
\urldef\tempurl%
\url{https://doi.org/10.2337/dc11-2434}
\showDOI{\tempurl}


\bibitem[Geurts et~al\mbox{.}(2019)]%
        {geurts_walkwithme_2019}
\bibfield{author}{\bibinfo{person}{Eva Geurts}, \bibinfo{person}{Fanny Van~Geel}, \bibinfo{person}{Peter Feys}, {and} \bibinfo{person}{Karin Coninx}.} \bibinfo{year}{2019}\natexlab{}.
\newblock \showarticletitle{{WalkWithMe}: {Personalized} {Goal} {Setting} and {Coaching} for {Walking} in {People} with {Multiple} {Sclerosis}}. In \bibinfo{booktitle}{\emph{Proceedings of the 27th {ACM} {Conference} on {User} {Modeling}, {Adaptation} and {Personalization}}} \emph{(\bibinfo{series}{{UMAP} '19})}. \bibinfo{publisher}{Association for Computing Machinery}, \bibinfo{address}{New York, NY, USA}, \bibinfo{pages}{51--60}.
\newblock
\showISBNx{978-1-4503-6021-0}
\urldef\tempurl%
\url{https://doi.org/10.1145/3320435.3320459}
\showDOI{\tempurl}


\bibitem[Gray et~al\mbox{.}(2018)]%
        {gray_barriers_2018}
\bibfield{author}{\bibinfo{person}{Wendy~N Gray}, \bibinfo{person}{Megan~R Schaefer}, \bibinfo{person}{Alana Resmini-Rawlinson}, {and} \bibinfo{person}{Scott~T Wagoner}.} \bibinfo{year}{2018}\natexlab{}.
\newblock \showarticletitle{Barriers to {Transition} {From} {Pediatric} to {Adult} {Care}: {A} {Systematic} {Review}}.
\newblock \bibinfo{journal}{\emph{Journal of Pediatric Psychology}} \bibinfo{volume}{43}, \bibinfo{number}{5} (\bibinfo{date}{June} \bibinfo{year}{2018}), \bibinfo{pages}{488--502}.
\newblock
\showISSN{0146-8693}
\urldef\tempurl%
\url{https://doi.org/10.1093/jpepsy/jsx142}
\showDOI{\tempurl}


\bibitem[Grimes et~al\mbox{.}(2009)]%
        {grimes_toward_2009}
\bibfield{author}{\bibinfo{person}{Andrea Grimes}, \bibinfo{person}{Desney Tan}, {and} \bibinfo{person}{Dan Morris}.} \bibinfo{year}{2009}\natexlab{}.
\newblock \showarticletitle{Toward technologies that support family reflections on health}. In \bibinfo{booktitle}{\emph{Proceedings of the 2009 {ACM} {International} {Conference} on {Supporting} {Group} {Work}}} \emph{(\bibinfo{series}{{GROUP} '09})}. \bibinfo{publisher}{Association for Computing Machinery}, \bibinfo{address}{New York, NY, USA}, \bibinfo{pages}{311--320}.
\newblock
\showISBNx{978-1-60558-500-0}
\urldef\tempurl%
\url{https://doi.org/10.1145/1531674.1531721}
\showDOI{\tempurl}


\bibitem[Heath et~al\mbox{.}(2017)]%
        {heath_parenting_2017}
\bibfield{author}{\bibinfo{person}{Gemma Heath}, \bibinfo{person}{Albert Farre}, {and} \bibinfo{person}{Karen Shaw}.} \bibinfo{year}{2017}\natexlab{}.
\newblock \showarticletitle{Parenting a child with chronic illness as they transition into adulthood: {A} systematic review and thematic synthesis of parents’ experiences}.
\newblock \bibinfo{journal}{\emph{Patient Education and Counseling}} \bibinfo{volume}{100}, \bibinfo{number}{1} (\bibinfo{date}{Jan.} \bibinfo{year}{2017}), \bibinfo{pages}{76--92}.
\newblock
\showISSN{0738-3991}
\urldef\tempurl%
\url{https://doi.org/10.1016/j.pec.2016.08.011}
\showDOI{\tempurl}


\bibitem[Holtz and Kanthawala(2020)]%
        {holtz_t1dlookslikeme_2020}
\bibfield{author}{\bibinfo{person}{Bree~E. Holtz} {and} \bibinfo{person}{Shaheen Kanthawala}.} \bibinfo{year}{2020}\natexlab{}.
\newblock \showarticletitle{\#{T1DLooksLikeMe}: {Exploring} {Self}-{Disclosure}, {Social} {Support}, and {Type} 1 {Diabetes} on {Instagram}}.
\newblock \bibinfo{journal}{\emph{Frontiers in Communication}}  \bibinfo{volume}{5} (\bibinfo{year}{2020}).
\newblock
\showISSN{2297-900X}
\urldef\tempurl%
\url{https://www.frontiersin.org/articles/10.3389/fcomm.2020.510278}
\showURL{%
\tempurl}


\bibitem[Hong et~al\mbox{.}(2020)]%
        {hong_using_2020}
\bibfield{author}{\bibinfo{person}{Matthew~K. Hong}, \bibinfo{person}{Udaya Lakshmi}, \bibinfo{person}{Kimberly Do}, \bibinfo{person}{Sampath Prahalad}, \bibinfo{person}{Thomas Olson}, \bibinfo{person}{Rosa~I. Arriaga}, {and} \bibinfo{person}{Lauren Wilcox}.} \bibinfo{year}{2020}\natexlab{}.
\newblock \showarticletitle{Using {Diaries} to {Probe} the {Illness} {Experiences} of {Adolescent} {Patients} and {Parental} {Caregivers}}. In \bibinfo{booktitle}{\emph{Proceedings of the 2020 {CHI} {Conference} on {Human} {Factors} in {Computing} {Systems}}}. \bibinfo{publisher}{ACM}, \bibinfo{address}{Honolulu HI USA}, \bibinfo{pages}{1--16}.
\newblock
\showISBNx{978-1-4503-6708-0}
\urldef\tempurl%
\url{https://doi.org/10.1145/3313831.3376426}
\showDOI{\tempurl}


\bibitem[Hong et~al\mbox{.}(2017)]%
        {hong_adolescent_2017}
\bibfield{author}{\bibinfo{person}{Matthew~K. Hong}, \bibinfo{person}{Lauren Wilcox}, \bibinfo{person}{Clayton Feustel}, \bibinfo{person}{Karen Wasileski-Masker}, \bibinfo{person}{Thomas~A. Olson}, {and} \bibinfo{person}{Stephen~F. Simoneaux}.} \bibinfo{year}{2017}\natexlab{}.
\newblock \showarticletitle{Adolescent and {Caregiver} use of a {Tethered} {Personal} {Health} {Record} {System}}.
\newblock \bibinfo{journal}{\emph{AMIA Annual Symposium Proceedings}}  \bibinfo{volume}{2016} (\bibinfo{date}{Feb.} \bibinfo{year}{2017}), \bibinfo{pages}{628--637}.
\newblock
\showISSN{1942-597X}
\urldef\tempurl%
\url{https://www.ncbi.nlm.nih.gov/pmc/articles/PMC5333234/}
\showURL{%
\tempurl}


\bibitem[Hong et~al\mbox{.}(2016)]%
        {hong_care_2016}
\bibfield{author}{\bibinfo{person}{Matthew~K. Hong}, \bibinfo{person}{Lauren Wilcox}, \bibinfo{person}{Daniel Machado}, \bibinfo{person}{Thomas~A. Olson}, {and} \bibinfo{person}{Stephen~F. Simoneaux}.} \bibinfo{year}{2016}\natexlab{}.
\newblock \showarticletitle{Care {Partnerships}: {Toward} {Technology} to {Support} {Teens}' {Participation} in {Their} {Health} {Care}}. In \bibinfo{booktitle}{\emph{Proceedings of the 2016 {CHI} {Conference} on {Human} {Factors} in {Computing} {Systems}}} \emph{(\bibinfo{series}{{CHI} '16})}. \bibinfo{publisher}{Association for Computing Machinery}, \bibinfo{address}{New York, NY, USA}, \bibinfo{pages}{5337--5349}.
\newblock
\showISBNx{978-1-4503-3362-7}
\urldef\tempurl%
\url{https://doi.org/10.1145/2858036.2858508}
\showDOI{\tempurl}


\bibitem[Huang et~al\mbox{.}(2014)]%
        {huang_preparing_2014}
\bibfield{author}{\bibinfo{person}{Jeannie~S. Huang}, \bibinfo{person}{Laura Terrones}, \bibinfo{person}{Trevor Tompane}, \bibinfo{person}{Lindsay Dillon}, \bibinfo{person}{Mark Pian}, \bibinfo{person}{Michael Gottschalk}, \bibinfo{person}{Gregory~J. Norman}, {and} \bibinfo{person}{L.~Kay Bartholomew}.} \bibinfo{year}{2014}\natexlab{}.
\newblock \showarticletitle{Preparing adolescents with chronic disease for transition to adult care: a technology program}.
\newblock \bibinfo{journal}{\emph{Pediatrics}} \bibinfo{volume}{133}, \bibinfo{number}{6} (\bibinfo{date}{June} \bibinfo{year}{2014}), \bibinfo{pages}{e1639--1646}.
\newblock
\showISSN{1098-4275}
\urldef\tempurl%
\url{https://doi.org/10.1542/peds.2013-2830}
\showDOI{\tempurl}


\bibitem[Huh et~al\mbox{.}(2014)]%
        {huh2014health}
\bibfield{author}{\bibinfo{person}{Jina Huh}, \bibinfo{person}{Leslie~S Liu}, \bibinfo{person}{Tina Neogi}, \bibinfo{person}{Kori Inkpen}, {and} \bibinfo{person}{Wanda Pratt}.} \bibinfo{year}{2014}\natexlab{}.
\newblock \showarticletitle{Health vlogs as social support for chronic illness management}.
\newblock \bibinfo{journal}{\emph{ACM Transactions on Computer-Human Interaction (TOCHI)}} \bibinfo{volume}{21}, \bibinfo{number}{4} (\bibinfo{year}{2014}), \bibinfo{pages}{1--31}.
\newblock


\bibitem[Huh-Yoo et~al\mbox{.}(2023)]%
        {huh-yoo_help_2023}
\bibfield{author}{\bibinfo{person}{Jina Huh-Yoo}, \bibinfo{person}{Afsaneh Razi}, \bibinfo{person}{Diep~N. Nguyen}, \bibinfo{person}{Sampada Regmi}, {and} \bibinfo{person}{Pamela~J. Wisniewski}.} \bibinfo{year}{2023}\natexlab{}.
\newblock \showarticletitle{“{Help} {Me}:” {Examining} {Youth}’s {Private} {Pleas} for {Support} and the {Responses} {Received} from {Peers} via {Instagram} {Direct} {Messages}}. In \bibinfo{booktitle}{\emph{Proceedings of the 2023 {CHI} {Conference} on {Human} {Factors} in {Computing} {Systems}}} \emph{(\bibinfo{series}{{CHI} '23})}. \bibinfo{publisher}{Association for Computing Machinery}, \bibinfo{address}{New York, NY, USA}, \bibinfo{pages}{1--14}.
\newblock
\showISBNx{978-1-4503-9421-5}
\urldef\tempurl%
\url{https://doi.org/10.1145/3544548.3581233}
\showDOI{\tempurl}


\bibitem[Hui et~al\mbox{.}(2012)]%
        {hui_mammibelli_2012}
\bibfield{author}{\bibinfo{person}{Mary Hui}, \bibinfo{person}{Christine Ly}, {and} \bibinfo{person}{Carman Neustaedter}.} \bibinfo{year}{2012}\natexlab{}.
\newblock \showarticletitle{{MammiBelli}: sharing baby activity levels between expectant mothers and their intimate social groups}. In \bibinfo{booktitle}{\emph{{CHI} '12 {Extended} {Abstracts} on {Human} {Factors} in {Computing} {Systems}}} \emph{(\bibinfo{series}{{CHI} {EA} '12})}. \bibinfo{publisher}{Association for Computing Machinery}, \bibinfo{address}{New York, NY, USA}, \bibinfo{pages}{1649--1654}.
\newblock
\showISBNx{978-1-4503-1016-1}
\urldef\tempurl%
\url{https://doi.org/10.1145/2212776.2223687}
\showDOI{\tempurl}


\bibitem[Jacobs et~al\mbox{.}(2019)]%
        {jacobs_i_2019}
\bibfield{author}{\bibinfo{person}{Maia Jacobs}, \bibinfo{person}{Galina Gheihman}, \bibinfo{person}{Krzysztof~Z. Gajos}, {and} \bibinfo{person}{Anoopum~S. Gupta}.} \bibinfo{year}{2019}\natexlab{}.
\newblock \showarticletitle{``{I} think we know more than our doctors'': {How} {Primary} {Caregivers} {Manage} {Care} {Teams} with {Limited} {Disease}-related {Expertise}}.
\newblock \bibinfo{journal}{\emph{Proceedings of the ACM on Human-Computer Interaction}} \bibinfo{volume}{3}, \bibinfo{number}{CSCW} (\bibinfo{date}{Nov.} \bibinfo{year}{2019}), \bibinfo{pages}{1--22}.
\newblock
\showISSN{2573-0142}
\urldef\tempurl%
\url{https://doi.org/10.1145/3359261}
\showDOI{\tempurl}


\bibitem[Jin et~al\mbox{.}(2023)]%
        {jin_understanding_2023}
\bibfield{author}{\bibinfo{person}{Yucheng Jin}, \bibinfo{person}{Wanling Cai}, \bibinfo{person}{Li Chen}, \bibinfo{person}{Yuwan Dai}, {and} \bibinfo{person}{Tonglin Jiang}.} \bibinfo{year}{2023}\natexlab{}.
\newblock \showarticletitle{Understanding {Disclosure} and {Support} for {Youth} {Mental} {Health} in {Social} {Music} {Communities}}.
\newblock \bibinfo{journal}{\emph{Proceedings of the ACM on Human-Computer Interaction}} \bibinfo{volume}{7}, \bibinfo{number}{CSCW1} (\bibinfo{date}{April} \bibinfo{year}{2023}), \bibinfo{pages}{153:1--153:32}.
\newblock
\urldef\tempurl%
\url{https://doi.org/10.1145/3579629}
\showDOI{\tempurl}


\bibitem[Jo et~al\mbox{.}(2020)]%
        {jo_mamas_2020}
\bibfield{author}{\bibinfo{person}{Eunkyung Jo}, \bibinfo{person}{Hyeonseok Bang}, \bibinfo{person}{Myeonghan Ryu}, \bibinfo{person}{Eun~Jee Sung}, \bibinfo{person}{Sungmook Leem}, {and} \bibinfo{person}{Hwajung Hong}.} \bibinfo{year}{2020}\natexlab{}.
\newblock \showarticletitle{{MAMAS}: {Supporting} {Parent}--{Child} {Mealtime} {Interactions} {Using} {Automated} {Tracking} and {Speech} {Recognition}}.
\newblock \bibinfo{journal}{\emph{Proceedings of the ACM on Human-Computer Interaction}} \bibinfo{volume}{4}, \bibinfo{number}{CSCW1} (\bibinfo{date}{May} \bibinfo{year}{2020}), \bibinfo{pages}{66:1--66:32}.
\newblock
\urldef\tempurl%
\url{https://doi.org/10.1145/3392876}
\showDOI{\tempurl}


\bibitem[Jo et~al\mbox{.}(2023)]%
        {jo_understanding_2023}
\bibfield{author}{\bibinfo{person}{Eunkyung Jo}, \bibinfo{person}{Daniel~A. Epstein}, \bibinfo{person}{Hyunhoon Jung}, {and} \bibinfo{person}{Young-Ho Kim}.} \bibinfo{year}{2023}\natexlab{}.
\newblock \showarticletitle{Understanding the {Benefits} and {Challenges} of {Deploying} {Conversational} {AI} {Leveraging} {Large} {Language} {Models} for {Public} {Health} {Intervention}}. In \bibinfo{booktitle}{\emph{Proceedings of the 2023 {CHI} {Conference} on {Human} {Factors} in {Computing} {Systems}}} \emph{(\bibinfo{series}{{CHI} '23})}. \bibinfo{publisher}{Association for Computing Machinery}, \bibinfo{address}{New York, NY, USA}, \bibinfo{pages}{1--16}.
\newblock
\showISBNx{978-1-4503-9421-5}
\urldef\tempurl%
\url{https://doi.org/10.1145/3544548.3581503}
\showDOI{\tempurl}


\bibitem[Kabir and Wiese(2023)]%
        {kabir_meta-synthesis_2023}
\bibfield{author}{\bibinfo{person}{Kazi~Sinthia Kabir} {and} \bibinfo{person}{Jason Wiese}.} \bibinfo{year}{2023}\natexlab{}.
\newblock \showarticletitle{A {Meta}-{Synthesis} of the {Barriers} and {Facilitators} for {Personal} {Informatics} {Systems}}.
\newblock \bibinfo{journal}{\emph{Proceedings of the ACM on Interactive, Mobile, Wearable and Ubiquitous Technologies}} \bibinfo{volume}{7}, \bibinfo{number}{3} (\bibinfo{date}{Sept.} \bibinfo{year}{2023}), \bibinfo{pages}{103:1--103:35}.
\newblock
\urldef\tempurl%
\url{https://doi.org/10.1145/3610893}
\showDOI{\tempurl}


\bibitem[Karkar et~al\mbox{.}(2017)]%
        {karkar_tummytrials_2017}
\bibfield{author}{\bibinfo{person}{Ravi Karkar}, \bibinfo{person}{Jessica Schroeder}, \bibinfo{person}{Daniel~A. Epstein}, \bibinfo{person}{Laura~R. Pina}, \bibinfo{person}{Jeffrey Scofield}, \bibinfo{person}{James Fogarty}, \bibinfo{person}{Julie~A. Kientz}, \bibinfo{person}{Sean~A. Munson}, \bibinfo{person}{Roger Vilardaga}, {and} \bibinfo{person}{Jasmine Zia}.} \bibinfo{year}{2017}\natexlab{}.
\newblock \showarticletitle{{TummyTrials}: {A} {Feasibility} {Study} of {Using} {Self}-{Experimentation} to {Detect} {Individualized} {Food} {Triggers}}. In \bibinfo{booktitle}{\emph{Proceedings of the 2017 {CHI} {Conference} on {Human} {Factors} in {Computing} {Systems}}} \emph{(\bibinfo{series}{{CHI} '17})}. \bibinfo{publisher}{Association for Computing Machinery}, \bibinfo{address}{New York, NY, USA}, \bibinfo{pages}{6850--6863}.
\newblock
\showISBNx{978-1-4503-4655-9}
\urldef\tempurl%
\url{https://doi.org/10.1145/3025453.3025480}
\showDOI{\tempurl}


\bibitem[Kaushansky et~al\mbox{.}(2017)]%
        {kaushansky_living_2017}
\bibfield{author}{\bibinfo{person}{Daniel Kaushansky}, \bibinfo{person}{Jarad Cox}, \bibinfo{person}{Chaka Dodson}, \bibinfo{person}{Miles McNeeley}, \bibinfo{person}{Sinthu Kumar}, {and} \bibinfo{person}{Ellen Iverson}.} \bibinfo{year}{2017}\natexlab{}.
\newblock \showarticletitle{Living a secret: {Disclosure} among adolescents and young adults with chronic illnesses}.
\newblock \bibinfo{journal}{\emph{Chronic Illness}} \bibinfo{volume}{13}, \bibinfo{number}{1} (\bibinfo{date}{March} \bibinfo{year}{2017}), \bibinfo{pages}{49--61}.
\newblock
\showISSN{1745-9206}
\urldef\tempurl%
\url{https://doi.org/10.1177/1742395316655855}
\showDOI{\tempurl}


\bibitem[Kawas et~al\mbox{.}(2020)]%
        {kawas_another_2020}
\bibfield{author}{\bibinfo{person}{Saba Kawas}, \bibinfo{person}{Ye Yuan}, \bibinfo{person}{Akeiylah DeWitt}, \bibinfo{person}{Qiao Jin}, \bibinfo{person}{Susanne Kirchner}, \bibinfo{person}{Abigail Bilger}, \bibinfo{person}{Ethan Grantham}, \bibinfo{person}{Julie~A Kientz}, \bibinfo{person}{Andrea Tartaro}, {and} \bibinfo{person}{Svetlana Yarosh}.} \bibinfo{year}{2020}\natexlab{}.
\newblock \showarticletitle{Another decade of {IDC} research: examining and reflecting on values and ethics}. In \bibinfo{booktitle}{\emph{Proceedings of the {Interaction} {Design} and {Children} {Conference}}} \emph{(\bibinfo{series}{{IDC} '20})}. \bibinfo{publisher}{Association for Computing Machinery}, \bibinfo{address}{New York, NY, USA}, \bibinfo{pages}{205--215}.
\newblock
\showISBNx{978-1-4503-7981-6}
\urldef\tempurl%
\url{https://doi.org/10.1145/3392063.3394436}
\showDOI{\tempurl}


\bibitem[Kaziunas et~al\mbox{.}(2017)]%
        {kaziunas_caring_2017}
\bibfield{author}{\bibinfo{person}{Elizabeth Kaziunas}, \bibinfo{person}{Mark~S. Ackerman}, \bibinfo{person}{Silvia Lindtner}, {and} \bibinfo{person}{Joyce~M. Lee}.} \bibinfo{year}{2017}\natexlab{}.
\newblock \showarticletitle{Caring through data: {Attending} to the social and emotional experiences of health datafication}.
\newblock \bibinfo{journal}{\emph{Proceedings of the 2017 ACM Conference on Computer Supported Cooperative Work and Social Computing}} (\bibinfo{year}{2017}).
\newblock
\urldef\tempurl%
\url{https://dl.acm.org/doi/abs/10.1145/2998181.2998303}
\showURL{%
\tempurl}


\bibitem[Kim et~al\mbox{.}(2019)]%
        {kim_toward_2019}
\bibfield{author}{\bibinfo{person}{Sung-In Kim}, \bibinfo{person}{Eunkyung Jo}, \bibinfo{person}{Myeonghan Ryu}, \bibinfo{person}{Inha Cha}, \bibinfo{person}{Young-Ho Kim}, \bibinfo{person}{Heejung Yoo}, {and} \bibinfo{person}{Hwajung Hong}.} \bibinfo{year}{2019}\natexlab{}.
\newblock \showarticletitle{Toward {Becoming} a {Better} {Self}: {Understanding} {Self}-{Tracking} {Experiences} of {Adolescents} with {Autism} {Spectrum} {Disorder} {Using} {Custom} {Trackers}}. In \bibinfo{booktitle}{\emph{Proceedings of the 13th {EAI} {International} {Conference} on {Pervasive} {Computing} {Technologies} for {Healthcare}}} \emph{(\bibinfo{series}{{PervasiveHealth}'19})}. \bibinfo{publisher}{Association for Computing Machinery}, \bibinfo{address}{New York, NY, USA}, \bibinfo{pages}{169--178}.
\newblock
\showISBNx{978-1-4503-6126-2}
\urldef\tempurl%
\url{https://doi.org/10.1145/3329189.3329209}
\showDOI{\tempurl}


\bibitem[Lebrun-Harris et~al\mbox{.}(2018)]%
        {lebrun-harris_transition_2018}
\bibfield{author}{\bibinfo{person}{Lydie~A. Lebrun-Harris}, \bibinfo{person}{Margaret~A. McManus}, \bibinfo{person}{Samhita~M. Ilango}, \bibinfo{person}{Mallory Cyr}, \bibinfo{person}{Sarah~Beth McLellan}, \bibinfo{person}{Marie~Y. Mann}, {and} \bibinfo{person}{Patience~H. White}.} \bibinfo{year}{2018}\natexlab{}.
\newblock \showarticletitle{Transition {Planning} {Among} {US} {Youth} {With} and {Without} {Special} {Health} {Care} {Needs}}.
\newblock \bibinfo{journal}{\emph{Pediatrics}} \bibinfo{volume}{142}, \bibinfo{number}{4} (\bibinfo{date}{Oct.} \bibinfo{year}{2018}), \bibinfo{pages}{e20180194}.
\newblock
\showISSN{0031-4005}
\urldef\tempurl%
\url{https://doi.org/10.1542/peds.2018-0194}
\showDOI{\tempurl}


\bibitem[Lee et~al\mbox{.}(2010)]%
        {lee_asthmon_2010}
\bibfield{author}{\bibinfo{person}{Hee~Rin Lee}, \bibinfo{person}{Wassa~R. Panont}, \bibinfo{person}{Brian Plattenburg}, \bibinfo{person}{Jean-Pierre De~La~Croix}, \bibinfo{person}{Dilip Patharachalam}, {and} \bibinfo{person}{Gregory Abowd}.} \bibinfo{year}{2010}\natexlab{}.
\newblock \showarticletitle{Asthmon: empowering asthmatic children's self-management with a virtual pet}. In \bibinfo{booktitle}{\emph{{CHI} '10 {Extended} {Abstracts} on {Human} {Factors} in {Computing} {Systems}}}. \bibinfo{publisher}{ACM}, \bibinfo{address}{Atlanta Georgia USA}, \bibinfo{pages}{3583--3588}.
\newblock
\showISBNx{978-1-60558-930-5}
\urldef\tempurl%
\url{https://doi.org/10.1145/1753846.1754022}
\showDOI{\tempurl}


\bibitem[Lemke et~al\mbox{.}(2018)]%
        {lemke_perceptions_2018}
\bibfield{author}{\bibinfo{person}{Monika Lemke}, \bibinfo{person}{Rachel Kappel}, \bibinfo{person}{Robert McCarter}, \bibinfo{person}{Lawrence D’Angelo}, {and} \bibinfo{person}{Lisa~K. Tuchman}.} \bibinfo{year}{2018}\natexlab{}.
\newblock \showarticletitle{Perceptions of {Health} {Care} {Transition} {Care} {Coordination} in {Patients} {With} {Chronic} {Illness}}.
\newblock \bibinfo{journal}{\emph{Pediatrics}} \bibinfo{volume}{141}, \bibinfo{number}{5} (\bibinfo{date}{May} \bibinfo{year}{2018}), \bibinfo{pages}{e20173168}.
\newblock
\showISSN{0031-4005}
\urldef\tempurl%
\url{https://doi.org/10.1542/peds.2017-3168}
\showDOI{\tempurl}


\bibitem[Lerch and Thrane(2019)]%
        {lerch_adolescents_2019}
\bibfield{author}{\bibinfo{person}{Matthew~F. Lerch} {and} \bibinfo{person}{Susan~E. Thrane}.} \bibinfo{year}{2019}\natexlab{}.
\newblock \showarticletitle{Adolescents with chronic illness and the transition to self-management: {A} systematic review}.
\newblock \bibinfo{journal}{\emph{Journal of Adolescence}}  \bibinfo{volume}{72} (\bibinfo{date}{April} \bibinfo{year}{2019}), \bibinfo{pages}{152--161}.
\newblock
\showISSN{1095-9254}
\urldef\tempurl%
\url{https://doi.org/10.1016/j.adolescence.2019.02.010}
\showDOI{\tempurl}


\bibitem[Li et~al\mbox{.}(2010)]%
        {li_stage-based_2010}
\bibfield{author}{\bibinfo{person}{Ian Li}, \bibinfo{person}{Anind Dey}, {and} \bibinfo{person}{Jodi Forlizzi}.} \bibinfo{year}{2010}\natexlab{}.
\newblock \showarticletitle{A stage-based model of personal informatics systems}. In \bibinfo{booktitle}{\emph{Proceedings of the {SIGCHI} {Conference} on {Human} {Factors} in {Computing} {Systems}}} \emph{(\bibinfo{series}{{CHI} '10})}. \bibinfo{publisher}{Association for Computing Machinery}, \bibinfo{address}{New York, NY, USA}, \bibinfo{pages}{557--566}.
\newblock
\showISBNx{978-1-60558-929-9}
\urldef\tempurl%
\url{https://doi.org/10.1145/1753326.1753409}
\showDOI{\tempurl}


\bibitem[Li et~al\mbox{.}(2020)]%
        {li_supporting_2020}
\bibfield{author}{\bibinfo{person}{Qingyang Li}, \bibinfo{person}{Clara Caldeira}, \bibinfo{person}{Daniel~A. Epstein}, {and} \bibinfo{person}{Yunan Chen}.} \bibinfo{year}{2020}\natexlab{}.
\newblock \showarticletitle{Supporting {Caring} among {Intergenerational} {Family} {Members} through {Family} {Fitness} {Tracking}}.
\newblock In \bibinfo{booktitle}{\emph{Proceedings of the 14th {EAI} {International} {Conference} on {Pervasive} {Computing} {Technologies} for {Healthcare}}}. \bibinfo{publisher}{Association for Computing Machinery}, \bibinfo{address}{New York, NY, USA}, \bibinfo{pages}{1--10}.
\newblock
\showISBNx{978-1-4503-7532-0}
\urldef\tempurl%
\url{https://doi.org/10.1145/3421937.3422018}
\showURL{%
\tempurl}


\bibitem[Lindsay et~al\mbox{.}(2011)]%
        {lindsay_barriers_2011}
\bibfield{author}{\bibinfo{person}{Sally Lindsay}, \bibinfo{person}{Shauna Kingsnorth}, {and} \bibinfo{person}{Yani Hamdani}.} \bibinfo{year}{2011}\natexlab{}.
\newblock \showarticletitle{Barriers and facilitators of chronic illness self-management among adolescents: a review and future directions}.
\newblock \bibinfo{journal}{\emph{Journal of Nursing and Healthcare of Chronic Illness}} \bibinfo{volume}{3}, \bibinfo{number}{3} (\bibinfo{year}{2011}), \bibinfo{pages}{186--208}.
\newblock
\showISSN{1752-9824}
\urldef\tempurl%
\url{https://doi.org/10.1111/j.1752-9824.2011.01090.x}
\showDOI{\tempurl}


\bibitem[Lukoff et~al\mbox{.}(2018)]%
        {lukoff_tablechat_2018}
\bibfield{author}{\bibinfo{person}{Kai Lukoff}, \bibinfo{person}{Taoxi Li}, \bibinfo{person}{Yuan Zhuang}, {and} \bibinfo{person}{Brian~Y. Lim}.} \bibinfo{year}{2018}\natexlab{}.
\newblock \showarticletitle{{TableChat}: {Mobile} {Food} {Journaling} to {Facilitate} {Family} {Support} for {Healthy} {Eating}}.
\newblock \bibinfo{journal}{\emph{Proceedings of the ACM on Human-Computer Interaction}} \bibinfo{volume}{2}, \bibinfo{number}{CSCW} (\bibinfo{date}{Nov.} \bibinfo{year}{2018}), \bibinfo{pages}{114:1--114:28}.
\newblock
\urldef\tempurl%
\url{https://doi.org/10.1145/3274383}
\showDOI{\tempurl}


\bibitem[Mamykina et~al\mbox{.}(2017)]%
        {mamykina_personal_2017}
\bibfield{author}{\bibinfo{person}{Lena Mamykina}, \bibinfo{person}{Elizabeth~M. Heitkemper}, \bibinfo{person}{Arlene~M. Smaldone}, \bibinfo{person}{Rita Kukafka}, \bibinfo{person}{Heather~J. Cole-Lewis}, \bibinfo{person}{Patricia~G. Davidson}, \bibinfo{person}{Elizabeth~D. Mynatt}, \bibinfo{person}{Andrea Cassells}, \bibinfo{person}{Jonathan~N. Tobin}, {and} \bibinfo{person}{George Hripcsak}.} \bibinfo{year}{2017}\natexlab{}.
\newblock \showarticletitle{Personal discovery in diabetes self-management: {Discovering} cause and effect using self-monitoring data}.
\newblock \bibinfo{journal}{\emph{Journal of Biomedical Informatics}}  \bibinfo{volume}{76} (\bibinfo{date}{Dec.} \bibinfo{year}{2017}), \bibinfo{pages}{1--8}.
\newblock
\showISSN{1532-0464}
\urldef\tempurl%
\url{https://doi.org/10.1016/j.jbi.2017.09.013}
\showDOI{\tempurl}


\bibitem[Mamykina et~al\mbox{.}(2006)]%
        {mamykina_investigating_2006}
\bibfield{author}{\bibinfo{person}{Lena Mamykina}, \bibinfo{person}{Elizabeth~D. Mynatt}, {and} \bibinfo{person}{David~R. Kaufman}.} \bibinfo{year}{2006}\natexlab{}.
\newblock \showarticletitle{Investigating health management practices of individuals with diabetes}. In \bibinfo{booktitle}{\emph{Proceedings of the {SIGCHI} {Conference} on {Human} {Factors} in {Computing} {Systems}}} \emph{(\bibinfo{series}{{CHI} '06})}. \bibinfo{publisher}{Association for Computing Machinery}, \bibinfo{address}{New York, NY, USA}, \bibinfo{pages}{927--936}.
\newblock
\showISBNx{978-1-59593-372-0}
\urldef\tempurl%
\url{https://doi.org/10.1145/1124772.1124910}
\showDOI{\tempurl}


\bibitem[McInroy(2016)]%
        {mcinroy_pitfalls_2016}
\bibfield{author}{\bibinfo{person}{Lauren~B. McInroy}.} \bibinfo{year}{2016}\natexlab{}.
\newblock \showarticletitle{Pitfalls, {Potentials}, and {Ethics} of {Online} {Survey} {Research}: {LGBTQ} and {Other} {Marginalized} and {Hard}-to-{Access} {Youths}}.
\newblock \bibinfo{journal}{\emph{Social Work Research}} \bibinfo{volume}{40}, \bibinfo{number}{2} (\bibinfo{date}{June} \bibinfo{year}{2016}), \bibinfo{pages}{83--94}.
\newblock
\showISSN{1070-5309}
\urldef\tempurl%
\url{https://doi.org/10.1093/swr/svw005}
\showDOI{\tempurl}


\bibitem[McKillop et~al\mbox{.}(2018)]%
        {mckillop_designing_2018}
\bibfield{author}{\bibinfo{person}{Mollie McKillop}, \bibinfo{person}{Lena Mamykina}, {and} \bibinfo{person}{Noémie Elhadad}.} \bibinfo{year}{2018}\natexlab{}.
\newblock \showarticletitle{Designing in the {Dark}: {Eliciting} {Self}-tracking {Dimensions} for {Understanding} {Enigmatic} {Disease}}. In \bibinfo{booktitle}{\emph{Proceedings of the 2018 {CHI} {Conference} on {Human} {Factors} in {Computing} {Systems}}} \emph{(\bibinfo{series}{{CHI} '18})}. \bibinfo{publisher}{Association for Computing Machinery}, \bibinfo{address}{New York, NY, USA}, \bibinfo{pages}{1--15}.
\newblock
\showISBNx{978-1-4503-5620-6}
\urldef\tempurl%
\url{https://doi.org/10.1145/3173574.3174139}
\showDOI{\tempurl}


\bibitem[Mentis et~al\mbox{.}(2017)]%
        {mentis_crafting_2017}
\bibfield{author}{\bibinfo{person}{Helena~M. Mentis}, \bibinfo{person}{Anita Komlodi}, \bibinfo{person}{Katrina Schrader}, \bibinfo{person}{Michael Phipps}, \bibinfo{person}{Ann Gruber-Baldini}, \bibinfo{person}{Karen Yarbrough}, {and} \bibinfo{person}{Lisa Shulman}.} \bibinfo{year}{2017}\natexlab{}.
\newblock \showarticletitle{Crafting a {View} of {Self}-{Tracking} {Data} in the {Clinical} {Visit}}. In \bibinfo{booktitle}{\emph{Proceedings of the 2017 {CHI} {Conference} on {Human} {Factors} in {Computing} {Systems}}} \emph{(\bibinfo{series}{{CHI} '17})}. \bibinfo{publisher}{Association for Computing Machinery}, \bibinfo{address}{New York, NY, USA}, \bibinfo{pages}{5800--5812}.
\newblock
\showISBNx{978-1-4503-4655-9}
\urldef\tempurl%
\url{https://doi.org/10.1145/3025453.3025589}
\showDOI{\tempurl}


\bibitem[Nikkhah et~al\mbox{.}(2022)]%
        {nikkhah_family_2022}
\bibfield{author}{\bibinfo{person}{Sarah Nikkhah}, \bibinfo{person}{Swaroop John}, \bibinfo{person}{Krishna~Supradeep Yalamarti}, \bibinfo{person}{Emily~L. Mueller}, {and} \bibinfo{person}{Andrew~D. Miller}.} \bibinfo{year}{2022}\natexlab{}.
\newblock \showarticletitle{Family {Care} {Coordination} in the {Children}'s {Hospital}: {Phases} and {Cycles} in the {Pediatric} {Cancer} {Caregiving} {Journey}}.
\newblock \bibinfo{journal}{\emph{Proceedings of the ACM on Human-Computer Interaction}} \bibinfo{volume}{6}, \bibinfo{number}{CSCW2} (\bibinfo{date}{Nov.} \bibinfo{year}{2022}), \bibinfo{pages}{296:1--296:30}.
\newblock
\urldef\tempurl%
\url{https://doi.org/10.1145/3555187}
\showDOI{\tempurl}


\bibitem[Nunes and Fitzpatrick(2018)]%
        {nunes_understanding_2018}
\bibfield{author}{\bibinfo{person}{Francisco Nunes} {and} \bibinfo{person}{Geraldine Fitzpatrick}.} \bibinfo{year}{2018}\natexlab{}.
\newblock \showarticletitle{Understanding the {Mundane} {Nature} of {Self}-care: {Ethnographic} {Accounts} of {People} {Living} with {Parkinson}'s}. In \bibinfo{booktitle}{\emph{Proceedings of the 2018 {CHI} {Conference} on {Human} {Factors} in {Computing} {Systems}}}. \bibinfo{publisher}{ACM}, \bibinfo{address}{Montreal QC Canada}, \bibinfo{pages}{1--15}.
\newblock
\showISBNx{978-1-4503-5620-6}
\urldef\tempurl%
\url{https://doi.org/10.1145/3173574.3173976}
\showDOI{\tempurl}


\bibitem[of~Pediatrics et~al\mbox{.}(2017)]%
        {aap2017bright}
\bibfield{author}{\bibinfo{person}{American~Academy of Pediatrics}, \bibinfo{person}{Jr Hagan, Joseph~F.}, \bibinfo{person}{Judith~S. Shaw}, {and} \bibinfo{person}{Paula~M. Duncan}.} \bibinfo{year}{2017}\natexlab{}.
\newblock \bibinfo{booktitle}{\emph{{Bright Futures Guidelines for Health Supervision of Infants, Children, and Adolescents}}}.
\newblock \bibinfo{publisher}{American Academy of Pediatrics}.
\newblock
\showISBNx{978-1-61002-022-0}
\urldef\tempurl%
\url{https://doi.org/10.1542/9781610020237}
\showDOI{\tempurl}


\bibitem[Park and Chen(2015)]%
        {park_individual_2015}
\bibfield{author}{\bibinfo{person}{Sun~Young Park} {and} \bibinfo{person}{Yunan Chen}.} \bibinfo{year}{2015}\natexlab{}.
\newblock \showarticletitle{Individual and {Social} {Recognition}: {Challenges} and {Opportunities} in {Migraine} {Management}}. In \bibinfo{booktitle}{\emph{Proceedings of the 18th {ACM} {Conference} on {Computer} {Supported} {Cooperative} {Work} \& {Social} {Computing}}} \emph{(\bibinfo{series}{{CSCW} '15})}. \bibinfo{publisher}{Association for Computing Machinery}, \bibinfo{address}{New York, NY, USA}, \bibinfo{pages}{1540--1551}.
\newblock
\showISBNx{978-1-4503-2922-4}
\urldef\tempurl%
\url{https://doi.org/10.1145/2675133.2675248}
\showDOI{\tempurl}


\bibitem[Parker et~al\mbox{.}(2019)]%
        {parker_2019_snowball}
\bibfield{author}{\bibinfo{person}{Charlie Parker}, \bibinfo{person}{Sam Scott}, {and} \bibinfo{person}{Alistair Geddes}.} \bibinfo{year}{2019}\natexlab{}.
\newblock \showarticletitle{Snowball Sampling}.
\newblock \bibinfo{journal}{\emph{SAGE research methods foundations}} (\bibinfo{year}{2019}).
\newblock


\bibitem[Pichon et~al\mbox{.}(2021)]%
        {pichon_divided_2021}
\bibfield{author}{\bibinfo{person}{Adrienne Pichon}, \bibinfo{person}{Kayla Schiffer}, \bibinfo{person}{Emma Horan}, \bibinfo{person}{Bria Massey}, \bibinfo{person}{Suzanne Bakken}, \bibinfo{person}{Lena Mamykina}, {and} \bibinfo{person}{Noemie Elhadad}.} \bibinfo{year}{2021}\natexlab{}.
\newblock \showarticletitle{Divided {We} {Stand}: {The} {Collaborative} {Work} of {Patients} and {Providers} in an {Enigmatic} {Chronic} {Disease}}.
\newblock \bibinfo{journal}{\emph{Proceedings of the ACM on Human-Computer Interaction}} \bibinfo{volume}{4}, \bibinfo{number}{CSCW3} (\bibinfo{date}{Jan.} \bibinfo{year}{2021}), \bibinfo{pages}{261:1--261:24}.
\newblock
\urldef\tempurl%
\url{https://doi.org/10.1145/3434170}
\showDOI{\tempurl}


\bibitem[Pina et~al\mbox{.}(2020)]%
        {pina_dreamcatcher_2020}
\bibfield{author}{\bibinfo{person}{Laura Pina}, \bibinfo{person}{Sang-Wha Sien}, \bibinfo{person}{Clarissa Song}, \bibinfo{person}{Teresa~M. Ward}, \bibinfo{person}{James Fogarty}, \bibinfo{person}{Sean~A. Munson}, {and} \bibinfo{person}{Julie~A. Kientz}.} \bibinfo{year}{2020}\natexlab{}.
\newblock \showarticletitle{{DreamCatcher}: {Exploring} {How} {Parents} and {School}-{Age} {Children} can {Track} and {Review} {Sleep} {Information} {Together}}.
\newblock \bibinfo{journal}{\emph{Proceedings of the ACM on Human-Computer Interaction}} \bibinfo{volume}{4}, \bibinfo{number}{CSCW1} (\bibinfo{date}{May} \bibinfo{year}{2020}), \bibinfo{pages}{70:1--70:25}.
\newblock
\urldef\tempurl%
\url{https://doi.org/10.1145/3392882}
\showDOI{\tempurl}


\bibitem[Pina et~al\mbox{.}(2017)]%
        {pina_personal_2017}
\bibfield{author}{\bibinfo{person}{Laura~R. Pina}, \bibinfo{person}{Sang-Wha Sien}, \bibinfo{person}{Teresa Ward}, \bibinfo{person}{Jason~C. Yip}, \bibinfo{person}{Sean~A. Munson}, \bibinfo{person}{James Fogarty}, {and} \bibinfo{person}{Julie~A. Kientz}.} \bibinfo{year}{2017}\natexlab{}.
\newblock \showarticletitle{From {Personal} {Informatics} to {Family} {Informatics}: {Understanding} {Family} {Practices} around {Health} {Monitoring}}. In \bibinfo{booktitle}{\emph{Proceedings of the 2017 {ACM} {Conference} on {Computer} {Supported} {Cooperative} {Work} and {Social} {Computing}}} \emph{(\bibinfo{series}{{CSCW} '17})}. \bibinfo{publisher}{Association for Computing Machinery}, \bibinfo{address}{New York, NY, USA}, \bibinfo{pages}{2300--2315}.
\newblock
\showISBNx{978-1-4503-4335-0}
\urldef\tempurl%
\url{https://doi.org/10.1145/2998181.2998362}
\showDOI{\tempurl}


\bibitem[Rahman et~al\mbox{.}(2021)]%
        {rahman_adolescentbot_2021}
\bibfield{author}{\bibinfo{person}{Rifat Rahman}, \bibinfo{person}{Md.~Rishadur Rahman}, \bibinfo{person}{Nafis~Irtiza Tripto}, \bibinfo{person}{Mohammed~Eunus Ali}, \bibinfo{person}{Sajid~Hasan Apon}, {and} \bibinfo{person}{Rifat Shahriyar}.} \bibinfo{year}{2021}\natexlab{}.
\newblock \showarticletitle{{AdolescentBot}: {Understanding} {Opportunities} for {Chatbots} in {Combating} {Adolescent} {Sexual} and {Reproductive} {Health} {Problems} in {Bangladesh}}. In \bibinfo{booktitle}{\emph{Proceedings of the 2021 {CHI} {Conference} on {Human} {Factors} in {Computing} {Systems}}} \emph{(\bibinfo{series}{{CHI} '21})}. \bibinfo{publisher}{Association for Computing Machinery}, \bibinfo{address}{New York, NY, USA}, \bibinfo{pages}{1--15}.
\newblock
\showISBNx{978-1-4503-8096-6}
\urldef\tempurl%
\url{https://doi.org/10.1145/3411764.3445694}
\showDOI{\tempurl}


\bibitem[Rotter(1966)]%
        {rotter_generalized_1966}
\bibfield{author}{\bibinfo{person}{Julian~B. Rotter}.} \bibinfo{year}{1966}\natexlab{}.
\newblock \showarticletitle{Generalized expectancies for internal versus external control of reinforcement}.
\newblock \bibinfo{journal}{\emph{Psychological Monographs: General and Applied}} \bibinfo{volume}{80}, \bibinfo{number}{1} (\bibinfo{year}{1966}), \bibinfo{pages}{1--28}.
\newblock
\showISSN{0096-9753}
\urldef\tempurl%
\url{https://doi.org/10.1037/h0092976}
\showDOI{\tempurl}
\newblock
\shownote{Place: US Publisher: American Psychological Association}.


\bibitem[Salinas(2023)]%
        {salinas_are_2023}
\bibfield{author}{\bibinfo{person}{Margaret~R. Salinas}.} \bibinfo{year}{2023}\natexlab{}.
\newblock \showarticletitle{Are {Your} {Participants} {Real}? {Dealing} with {Fraud} in {Recruiting} {Older} {Adults} {Online}}.
\newblock \bibinfo{journal}{\emph{Western Journal of Nursing Research}} \bibinfo{volume}{45}, \bibinfo{number}{1} (\bibinfo{date}{Jan.} \bibinfo{year}{2023}), \bibinfo{pages}{93--99}.
\newblock
\showISSN{0193-9459}
\urldef\tempurl%
\url{https://doi.org/10.1177/01939459221098468}
\showDOI{\tempurl}
\newblock
\shownote{Publisher: SAGE Publications Inc}.


\bibitem[Schroeder et~al\mbox{.}(2018)]%
        {schroeder_examining_2018}
\bibfield{author}{\bibinfo{person}{Jessica Schroeder}, \bibinfo{person}{Chia-Fang Chung}, \bibinfo{person}{Daniel~A. Epstein}, \bibinfo{person}{Ravi Karkar}, \bibinfo{person}{Adele Parsons}, \bibinfo{person}{Natalia Murinova}, \bibinfo{person}{James Fogarty}, {and} \bibinfo{person}{Sean~A. Munson}.} \bibinfo{year}{2018}\natexlab{}.
\newblock \showarticletitle{Examining {Self}-{Tracking} by {People} with {Migraine}: {Goals}, {Needs}, and {Opportunities} in a {Chronic} {Health} {Condition}}. In \bibinfo{booktitle}{\emph{Proceedings of the 2018 {Designing} {Interactive} {Systems} {Conference}}} \emph{(\bibinfo{series}{{DIS} '18})}. \bibinfo{publisher}{Association for Computing Machinery}, \bibinfo{address}{New York, NY, USA}, \bibinfo{pages}{135--148}.
\newblock
\showISBNx{978-1-4503-5198-0}
\urldef\tempurl%
\url{https://doi.org/10.1145/3196709.3196738}
\showDOI{\tempurl}


\bibitem[Schroeder et~al\mbox{.}(2020)]%
        {schroeder_examining_2020}
\bibfield{author}{\bibinfo{person}{Jessica Schroeder}, \bibinfo{person}{Ravi Karkar}, \bibinfo{person}{Natalia Murinova}, \bibinfo{person}{James Fogarty}, {and} \bibinfo{person}{Sean~A. Munson}.} \bibinfo{year}{2020}\natexlab{}.
\newblock \showarticletitle{Examining {Opportunities} for {Goal}-{Directed} {Self}-{Tracking} to {Support} {Chronic} {Condition} {Management}}.
\newblock \bibinfo{journal}{\emph{Proceedings of the ACM on Interactive, Mobile, Wearable and Ubiquitous Technologies}} \bibinfo{volume}{3}, \bibinfo{number}{4} (\bibinfo{date}{Sept.} \bibinfo{year}{2020}), \bibinfo{pages}{151:1--151:26}.
\newblock
\urldef\tempurl%
\url{https://doi.org/10.1145/3369809}
\showDOI{\tempurl}


\bibitem[Seo et~al\mbox{.}(2021)]%
        {seo_challenges_2021}
\bibfield{author}{\bibinfo{person}{Woosuk Seo}, \bibinfo{person}{Ayse~G. Buyuktur}, \bibinfo{person}{Sung~Won Choi}, \bibinfo{person}{Laura Sedig}, {and} \bibinfo{person}{Sun~Young Park}.} \bibinfo{year}{2021}\natexlab{}.
\newblock \showarticletitle{Challenges in the {Parent}-{Child} {Communication} of {Health}-related {Information} in {Pediatric} {Cancer} {Care}}.
\newblock \bibinfo{journal}{\emph{Proceedings of the ACM on Human-Computer Interaction}} \bibinfo{volume}{5}, \bibinfo{number}{CSCW1} (\bibinfo{date}{April} \bibinfo{year}{2021}), \bibinfo{pages}{110:1--110:24}.
\newblock
\urldef\tempurl%
\url{https://doi.org/10.1145/3449184}
\showDOI{\tempurl}


\bibitem[Shin and Holtz(2019)]%
        {shin_towards_2019}
\bibfield{author}{\bibinfo{person}{Ji~Youn Shin} {and} \bibinfo{person}{Bree~E. Holtz}.} \bibinfo{year}{2019}\natexlab{}.
\newblock \showarticletitle{Towards {Better} {Transitions} for {Children} with {Diabetes}: {User} {Experiences} on a {Mobile} {Health} {App}}. In \bibinfo{booktitle}{\emph{Proceedings of the 18th {ACM} {International} {Conference} on {Interaction} {Design} and {Children}}} \emph{(\bibinfo{series}{{IDC} '19})}. \bibinfo{publisher}{Association for Computing Machinery}, \bibinfo{address}{New York, NY, USA}, \bibinfo{pages}{623--628}.
\newblock
\showISBNx{978-1-4503-6690-8}
\urldef\tempurl%
\url{https://doi.org/10.1145/3311927.3325319}
\showDOI{\tempurl}


\bibitem[Shin et~al\mbox{.}(2022)]%
        {shin_more_2022}
\bibfield{author}{\bibinfo{person}{Ji~Youn Shin}, \bibinfo{person}{Wei Peng}, {and} \bibinfo{person}{Hee~Rin Lee}.} \bibinfo{year}{2022}\natexlab{}.
\newblock \showarticletitle{More than {Bedtime} and the {Bedroom}: {Sleep} {Management} as a {Collaborative} {Work} for the {Family}}. In \bibinfo{booktitle}{\emph{Proceedings of the 2022 {CHI} {Conference} on {Human} {Factors} in {Computing} {Systems}}} \emph{(\bibinfo{series}{{CHI} '22})}. \bibinfo{publisher}{Association for Computing Machinery}, \bibinfo{address}{New York, NY, USA}, \bibinfo{pages}{1--16}.
\newblock
\showISBNx{978-1-4503-9157-3}
\urldef\tempurl%
\url{https://doi.org/10.1145/3491102.3517535}
\showDOI{\tempurl}


\bibitem[Steptoe and Wardle(2001)]%
        {steptoe_locus_2001}
\bibfield{author}{\bibinfo{person}{Andrew Steptoe} {and} \bibinfo{person}{Jane Wardle}.} \bibinfo{year}{2001}\natexlab{}.
\newblock \showarticletitle{Locus of control and health behaviour revisited: {A} multivariate analysis of young adults from 18 countries}.
\newblock \bibinfo{journal}{\emph{British Journal of Psychology}} \bibinfo{volume}{92}, \bibinfo{number}{4} (\bibinfo{year}{2001}), \bibinfo{pages}{659--672}.
\newblock
\showISSN{2044-8295}
\urldef\tempurl%
\url{https://doi.org/10.1348/000712601162400}
\showDOI{\tempurl}


\bibitem[Su et~al\mbox{.}(2024)]%
        {su2024data}
\bibfield{author}{\bibinfo{person}{Zhaoyuan Su}, \bibinfo{person}{Yunan Chen}, {et~al\mbox{.}}} \bibinfo{year}{2024}\natexlab{}.
\newblock \showarticletitle{Data-Driven Technology for Children’s Health and Wellbeing: A Systematic Review}.
\newblock \bibinfo{journal}{\emph{Foundations and Trends{\textregistered} in Human-Computer Interaction}} \bibinfo{volume}{18}, \bibinfo{number}{1} (\bibinfo{year}{2024}), \bibinfo{pages}{1--99}.
\newblock


\bibitem[Tinschert et~al\mbox{.}(2017)]%
        {tinschert_potential_2017}
\bibfield{author}{\bibinfo{person}{Peter Tinschert}, \bibinfo{person}{Robert Jakob}, \bibinfo{person}{Filipe Barata}, \bibinfo{person}{Jan-Niklas Kramer}, {and} \bibinfo{person}{Tobias Kowatsch}.} \bibinfo{year}{2017}\natexlab{}.
\newblock \showarticletitle{The {Potential} of {Mobile} {Apps} for {Improving} {Asthma} {Self}-{Management}: {A} {Review} of {Publicly} {Available} and {Well}-{Adopted} {Asthma} {Apps}}.
\newblock \bibinfo{journal}{\emph{JMIR mHealth and uHealth}} \bibinfo{volume}{5}, \bibinfo{number}{8} (\bibinfo{date}{Aug.} \bibinfo{year}{2017}), \bibinfo{pages}{e7177}.
\newblock
\urldef\tempurl%
\url{https://doi.org/10.2196/mhealth.7177}
\showDOI{\tempurl}


\bibitem[Toscos et~al\mbox{.}(2012)]%
        {toscos_best_2012}
\bibfield{author}{\bibinfo{person}{Tammy Toscos}, \bibinfo{person}{Kay Connelly}, {and} \bibinfo{person}{Yvonne Rogers}.} \bibinfo{year}{2012}\natexlab{}.
\newblock \showarticletitle{Best intentions: health monitoring technology and children}. In \bibinfo{booktitle}{\emph{Proceedings of the {SIGCHI} {Conference} on {Human} {Factors} in {Computing} {Systems}}} \emph{(\bibinfo{series}{{CHI} '12})}. \bibinfo{publisher}{Association for Computing Machinery}, \bibinfo{address}{New York, NY, USA}, \bibinfo{pages}{1431--1440}.
\newblock
\showISBNx{978-1-4503-1015-4}
\urldef\tempurl%
\url{https://doi.org/10.1145/2207676.2208603}
\showDOI{\tempurl}


\bibitem[Turner and Kelly(2000)]%
        {turner_emotional_2000}
\bibfield{author}{\bibinfo{person}{Jane Turner} {and} \bibinfo{person}{Brian Kelly}.} \bibinfo{year}{2000}\natexlab{}.
\newblock \showarticletitle{Emotional dimensions of chronic disease}.
\newblock \bibinfo{journal}{\emph{Western Journal of Medicine}} \bibinfo{volume}{172}, \bibinfo{number}{2} (\bibinfo{date}{Feb.} \bibinfo{year}{2000}), \bibinfo{pages}{124--128}.
\newblock
\showISSN{0093-0415}
\urldef\tempurl%
\url{https://www.ncbi.nlm.nih.gov/pmc/articles/PMC1070773/}
\showURL{%
\tempurl}


\bibitem[van~der Velden and El~Emam(2013)]%
        {van_der_velden_not_2013}
\bibfield{author}{\bibinfo{person}{Maja van~der Velden} {and} \bibinfo{person}{Khaled El~Emam}.} \bibinfo{year}{2013}\natexlab{}.
\newblock \showarticletitle{“{Not} all my friends need to know”: a qualitative study of teenage patients, privacy, and social media}.
\newblock \bibinfo{journal}{\emph{Journal of the American Medical Informatics Association}} \bibinfo{volume}{20}, \bibinfo{number}{1} (\bibinfo{date}{Jan.} \bibinfo{year}{2013}), \bibinfo{pages}{16--24}.
\newblock
\showISSN{1067-5027}
\urldef\tempurl%
\url{https://doi.org/10.1136/amiajnl-2012-000949}
\showDOI{\tempurl}


\bibitem[Wang et~al\mbox{.}(2017)]%
        {wang_quantified_2017}
\bibfield{author}{\bibinfo{person}{Junqing Wang}, \bibinfo{person}{Aisling~Ann O'Kane}, \bibinfo{person}{Nikki Newhouse}, \bibinfo{person}{Geraint~Rhys Sethu-Jones}, {and} \bibinfo{person}{Kaya de Barbaro}.} \bibinfo{year}{2017}\natexlab{}.
\newblock \showarticletitle{Quantified {Baby}: {Parenting} and the {Use} of a {Baby} {Wearable} in the {Wild}}.
\newblock \bibinfo{journal}{\emph{Proceedings of the ACM on Human-Computer Interaction}} \bibinfo{volume}{1}, \bibinfo{number}{CSCW} (\bibinfo{date}{Dec.} \bibinfo{year}{2017}), \bibinfo{pages}{108:1--108:19}.
\newblock
\urldef\tempurl%
\url{https://doi.org/10.1145/3134743}
\showDOI{\tempurl}


\bibitem[Westeyn et~al\mbox{.}(2012)]%
        {westeyn_monitoring_2012}
\bibfield{author}{\bibinfo{person}{Tracy~L. Westeyn}, \bibinfo{person}{Gregory~D. Abowd}, \bibinfo{person}{Thad~E. Starner}, \bibinfo{person}{Jeremy~M. Johnson}, \bibinfo{person}{Peter~W. Presti}, {and} \bibinfo{person}{Kimberly~A. Weaver}.} \bibinfo{year}{2012}\natexlab{}.
\newblock \showarticletitle{Monitoring children’s developmental progress using augmented toys and activity recognition}.
\newblock \bibinfo{journal}{\emph{Personal and Ubiquitous Computing}} \bibinfo{volume}{16}, \bibinfo{number}{2} (\bibinfo{date}{Feb.} \bibinfo{year}{2012}), \bibinfo{pages}{169--191}.
\newblock
\showISSN{1617-4917}
\urldef\tempurl%
\url{https://doi.org/10.1007/s00779-011-0386-0}
\showDOI{\tempurl}


\bibitem[White et~al\mbox{.}(2018)]%
        {white_supporting_2018}
\bibfield{author}{\bibinfo{person}{Patience~H. White}, \bibinfo{person}{W.~Carl Cooley}, \bibinfo{person}{{Transitions Clinical Report Authoring Group}}, \bibinfo{person}{{American Academy Of Pediatrics}}, \bibinfo{person}{{American Academy Of Family Physicians}}, {and} \bibinfo{person}{{American College Of Physicians}}.} \bibinfo{year}{2018}\natexlab{}.
\newblock \showarticletitle{Supporting the {Health} {Care} {Transition} {From} {Adolescence} to {Adulthood} in the {Medical} {Home}}.
\newblock \bibinfo{journal}{\emph{Pediatrics}} \bibinfo{volume}{142}, \bibinfo{number}{5} (\bibinfo{date}{Nov.} \bibinfo{year}{2018}), \bibinfo{pages}{e20182587}.
\newblock
\showISSN{0031-4005}
\urldef\tempurl%
\url{https://doi.org/10.1542/peds.2018-2587}
\showDOI{\tempurl}


\bibitem[Whitman et~al\mbox{.}(2021)]%
        {whitman_bodily_2021}
\bibfield{author}{\bibinfo{person}{Samantha~A. Whitman}, \bibinfo{person}{Kathleen~H. Pine}, \bibinfo{person}{Bjorg Thorsteinsdottir}, \bibinfo{person}{Paige Organick-Lee}, \bibinfo{person}{Anjali Thota}, \bibinfo{person}{Nataly~R. Espinoza~Suarez}, \bibinfo{person}{Erik~W. Johnston}, {and} \bibinfo{person}{Kasey~R. Boehmer}.} \bibinfo{year}{2021}\natexlab{}.
\newblock \showarticletitle{Bodily {Experiences} of {Illness} and {Treatment} as {Information} {Work}: {The} {Case} of {Chronic} {Kidney} {Disease}}.
\newblock \bibinfo{journal}{\emph{Proceedings of the ACM on Human-Computer Interaction}} \bibinfo{volume}{5}, \bibinfo{number}{CSCW2} (\bibinfo{date}{Oct.} \bibinfo{year}{2021}), \bibinfo{pages}{383:1--383:28}.
\newblock
\urldef\tempurl%
\url{https://doi.org/10.1145/3479527}
\showDOI{\tempurl}


\bibitem[Williams and Moser(2019)]%
        {williams_2019_art}
\bibfield{author}{\bibinfo{person}{Michael Williams} {and} \bibinfo{person}{Tami Moser}.} \bibinfo{year}{2019}\natexlab{}.
\newblock \showarticletitle{The Art of Coding and Thematic Exploration in Qualitative Research}.
\newblock \bibinfo{journal}{\emph{International management review}} \bibinfo{volume}{15}, \bibinfo{number}{1} (\bibinfo{year}{2019}), \bibinfo{pages}{45--55}.
\newblock


\bibitem[Woolfall(2023)]%
        {woolfall_identifying_2023}
\bibfield{author}{\bibinfo{person}{Kerry Woolfall}.} \bibinfo{year}{2023}\natexlab{}.
\newblock \showarticletitle{Identifying and preventing fraudulent participation in qualitative research}.
\newblock \bibinfo{journal}{\emph{Archives of Disease in Childhood}} \bibinfo{volume}{108}, \bibinfo{number}{6} (\bibinfo{date}{June} \bibinfo{year}{2023}), \bibinfo{pages}{421--422}.
\newblock
\showISSN{0003-9888, 1468-2044}
\urldef\tempurl%
\url{https://doi.org/10.1136/archdischild-2023-325328}
\showDOI{\tempurl}


\bibitem[Wu et~al\mbox{.}(2023)]%
        {wu_ubi-asthma_2023}
\bibfield{author}{\bibinfo{person}{Yuan Wu}, \bibinfo{person}{Jian Zhang}, \bibinfo{person}{Yanjiao Chen}, \bibinfo{person}{Junkongshuai Wang}, \bibinfo{person}{Wuxuan Shi}, {and} \bibinfo{person}{Qian Zhang}.} \bibinfo{year}{2023}\natexlab{}.
\newblock \showarticletitle{Ubi-{Asthma}: {Toward} {Ubiquitous} {Asthma} {Detection} {Using} the {Smartwatch}}.
\newblock \bibinfo{journal}{\emph{IEEE Internet of Things Journal}} \bibinfo{volume}{10}, \bibinfo{number}{13} (\bibinfo{date}{July} \bibinfo{year}{2023}), \bibinfo{pages}{11576--11587}.
\newblock
\showISSN{2327-4662}
\urldef\tempurl%
\url{https://doi.org/10.1109/JIOT.2023.3243188}
\showDOI{\tempurl}


\bibitem[Yeo and Sawyer(2005)]%
        {yeo_chronic_2005}
\bibfield{author}{\bibinfo{person}{Michele Yeo} {and} \bibinfo{person}{Susan Sawyer}.} \bibinfo{year}{2005}\natexlab{}.
\newblock \showarticletitle{Chronic illness and disability}.
\newblock \bibinfo{journal}{\emph{BMJ}} \bibinfo{volume}{330}, \bibinfo{number}{7493} (\bibinfo{date}{March} \bibinfo{year}{2005}), \bibinfo{pages}{721--723}.
\newblock
\showISSN{0959-8138, 1756-1833}
\urldef\tempurl%
\url{https://doi.org/10.1136/bmj.330.7493.721}
\showDOI{\tempurl}


\bibitem[Yu and McDonald(2023)]%
        {yu_conflicts_2023}
\bibfield{author}{\bibinfo{person}{Yihan Yu} {and} \bibinfo{person}{David~W. McDonald}.} \bibinfo{year}{2023}\natexlab{}.
\newblock \showarticletitle{Conflicts of {Control}: {Continuous} {Blood} {Glucose} {Monitoring} and {Coordinated} {Caregiving} for {Teenagers} with {Type} 1 {Diabetes}}.
\newblock \bibinfo{journal}{\emph{Proceedings of the ACM on Human-Computer Interaction}} \bibinfo{volume}{7}, \bibinfo{number}{CSCW2} (\bibinfo{date}{Oct.} \bibinfo{year}{2023}), \bibinfo{pages}{306:1--306:32}.
\newblock
\urldef\tempurl%
\url{https://doi.org/10.1145/3610097}
\showDOI{\tempurl}


\bibitem[Zehrung and Chen(2023)]%
        {zehrung2023self}
\bibfield{author}{\bibinfo{person}{Rachael~F. Zehrung} {and} \bibinfo{person}{Yunan Chen}.} \bibinfo{year}{2023}\natexlab{}.
\newblock \showarticletitle{Self-Expression and Sharing around Chronic Illness on TikTok}. In \bibinfo{booktitle}{\emph{AMIA Annual Symposium Proceedings}}, Vol.~\bibinfo{volume}{2023}. American Medical Informatics Association, \bibinfo{pages}{1334}.
\newblock


\end{thebibliography}
\end{document}